\documentclass[aip, twocolumn, superscriptaddress, graphicx]{revtex4}
\usepackage{amssymb}
\usepackage{graphicx}
\usepackage{amsmath}
\usepackage{amsfonts}
\usepackage{epsfig}
\usepackage{epstopdf}
\usepackage{xcolor}
\usepackage[english]{babel}

\graphicspath{{ffunc/}{figures/}}
\draft

\begin{document}

\title{Thouless energy in Josephson SN-N-NS bridges}

\author{S.~V.~Bakurskiy}
\affiliation{Skobeltsyn Institute of Nuclear Physics, Lomonosov Moscow State University,
Moscow 119991, Russian Federation}
\affiliation{Dukhov All-Russia Research Institute of Automatics, Moscow 101000, Russia}
\author{V.~I.~Ruzhickiy}
\affiliation{Skobeltsyn Institute of Nuclear Physics, Lomonosov Moscow State University,
Moscow 119991, Russian Federation}
\affiliation{Dukhov All-Russia Research Institute of Automatics, Moscow 101000, Russia}
\author{A.~A.~Neilo}
\affiliation{Skobeltsyn Institute of Nuclear Physics, Lomonosov Moscow State University,
Moscow 119991, Russian Federation}
\author{N.~V.~Klenov}
\affiliation{Faculty of Physics, Lomonosov Moscow State University, Moscow 119992,
Russian Federation}
\affiliation{Dukhov All-Russia Research Institute of Automatics, Moscow 101000, Russia}
\author{I.~I.~Soloviev}
\affiliation{Skobeltsyn Institute of Nuclear Physics, Lomonosov Moscow State University,
Moscow 119991, Russian Federation}
\affiliation{Dukhov All-Russia Research Institute of Automatics, Moscow 101000, Russia}
\author{A.~A.~Elistratova}
\affiliation{Dukhov All-Russia Research Institute of Automatics, Moscow 101000, Russia}
\affiliation{Center for Advanced Mesoscience and Nanotechnology, Moscow Institute of Physics and Technology, 9 Institutskiy per., Dolgoprudny, 141700, Russia}
\author{A.~G.~Shishkin}
\affiliation{Dukhov All-Russia Research Institute of Automatics, Moscow 101000, Russia}
\affiliation{Center for Advanced Mesoscience and Nanotechnology, Moscow Institute of Physics and Technology, 9 Institutskiy per., Dolgoprudny, 141700, Russia}
\author{V.~S.~Stolyarov}
\affiliation{Dukhov All-Russia Research Institute of Automatics, Moscow 101000, Russia}
\affiliation{Center for Advanced Mesoscience and Nanotechnology, Moscow Institute of Physics and Technology, 9 Institutskiy per., Dolgoprudny, 141700, Russia}
\author{M.~Yu.~Kupriyanov}
\affiliation{Skobeltsyn Institute of Nuclear Physics, Lomonosov Moscow State University,
Moscow 119991, Russian Federation}

\date{\today }

\begin{abstract}
We have studied the Thouless energy in Josephson superconductor-normal metal-superconductor (SN-N-NS) bridges analytically and numerically, taking into account the influence of the sub-electrode regions. We have found a significant suppression of the Thouless energy with increasing interfacial resistance, in agreement with experimental results. The analysis of the temperature dependence of the critical current in Josephson junctions in comparison with the expressions for the Thouless energy may allow the determination of the interface parameters of S and N-layers. 
\end{abstract}

\maketitle

In recent years, there has been a reawakened interest in Josephson structures in which the weak coupling region has a metallic type of conductivity \cite{brinkman1999, Holmes,shelly2017weak,Tolp,semenov2019very,collins2022superconducting,thompson2022effects,tolpygo2023development,10431604,soloviev2021miniaturization,nano13121873}. 
Such structures are expected to overcome the limitations on the degree of integration that occur in superconducting devices for digital information processing.
The steady-state properties of SNS Josephson sandwiches and SN-N-NS bridges have been well studied \cite{likharev1979superconducting,golubov2004current,likharev1976sov,likharev1981boundary,zaikin1981theory,dubos2001josephson,hammer2007density,marychev2020josephson,soloviev2021miniaturization,nano13121873}.

\begin{figure*}[t]
	\centering
    \includegraphics[width=0.8\linewidth]{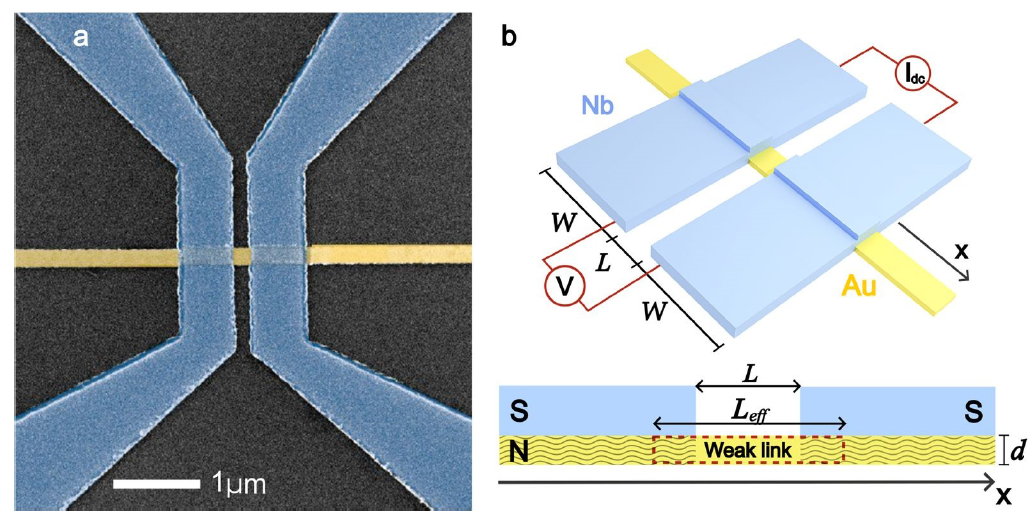}
	\caption{a) Scanning electron microscope image of an Nb/Au-Au-Au/Nb bridge with superconducting Nb electrodes and an Au
weak link. b) Top part: three-dimensional sketch of the Josephson superconductor/normal metal/superconductor (SN-N-NS) bridge.
The sample is connected using a four-point scheme. Bottom part: side view of the Josephson SN-N-NS bridge. Wavy lines indicate the
proximized region of the N-layer under the S-electrode, and the red frame indicates the effective weak link.}
	\label{Sketch}
\end{figure*}

In \cite{likharev1979superconducting}, assuming the fulfillment of rigid boundary conditions at the SN interfaces, it was shown that in the case when the distance, $L,$ between the S-electrodes significantly exceeds the characteristic decay length of superconducting correlations in the N-layer, $\xi=(D/2\pi T)^{1/2},$ the decay of the critical current, $I_c$ with increasing $L$ depends significantly on the ratio between the operating temperature, $T,$ and the critical temperature of the superconducting electrodes, $T_c$ ($D$ is the diffusion coefficient of an ordinary metal). At small temperature $(T\ll T_c)$ there is a power dependence ($I_{c}\propto 1/L^2$). With increasing temperature the dependence of $I_c(L)$ becomes exponential ($I_{c}\propto\exp\{-L/\xi\}$).  

In \cite{dubos2001josephson} these studies were complemented by a more detailed analysis of the temperature dependence of the critical current at low temperatures $T\ll T_c$. Numerical calculations carried out in \cite{dubos2001josephson} allowed the authors to propose an approximation formula for {$I_c$}:
\begin{equation}
\frac{eI_{c}R_{n}}{E_{T}}=\alpha \left( 1-\beta \exp \left\{-\frac{\alpha
E_{T}}{3.2T}\right\}\right) ,~E_{T}=\frac{D}{L^{2}},
\label{Ic(EthS)}
\end{equation}
where the coefficients $\alpha \approx10.82$ and $\beta \approx 1/3$ are {the fitting parameters} found in the limit of small Thouless energy, $E_{T},$ compared to the magnitude of the S-electrode order parameter, $\Delta$. In the opposite limit, $E_{T}\gg T,$ the dependence $I_{c} (E_{T})$ has the following form:
\begin{equation}
\frac{eI_{c}R_{n}}{E_{T}}=\frac{32}{3+2\sqrt{2}}\left( \frac{2\pi T}{E_{T}}%
\right) ^{3/2}\exp \left\{ -\sqrt{\frac{2\pi T}{E_{T}}}\right\} 
\label{Ic(EthL)}
\end{equation}

The publication of these results \cite{dubos2001josephson} stimulated the study of SNS-based structures. The aim was to experimentally determine the Thouless energy \cite{dubos2001josephson,hammer2007density,giazotto2006opportunities,Angers_2008,courtois2008origin,PhysRevB.78.052506,García_2009,giazotto2010superconducting,frielinghaus2010josephson,jung2011superconducting,batov2012critical3362877,PhysRevB.89.104507,pekola2015towards,paajaste2015pb,de2016interplay,jabdaraghi2016low,skryabina2017josephson,kim2017strong,shishkin2020planar,Zhang2020,murani2020long,skryabina2021environment,golod2021reconfigurable,sotnichuk2022long,babich2023limitations}. Experimental studies have shown that the values of the parameter $\alpha$ in the expression (\ref{Ic(EthS)}) obtained in \cite{dubos2001josephson} are at least several times larger than their experimental values. As a possible reason for this discrepancy, the presence of finite transparency of the SN boundaries of the studied Josephson contacts has been hypothesized. To account for this in ref. \cite{hammer2007density}, a renormalization of the Thouless energy was proposed by introducing a normalization coefficient: 
\begin{equation}
E_{T}^{eff}=E_{T}\frac{Ar^{B}}{C+r},~r=\frac{G_{N}}{G_{B}}.
\label{EthL_eff}
\end{equation}
Here $G_{N}$ and $G_{B}$ are 
the conductance of the normal wire  and the SN interface, respectively, $A,$ $B,$ and $C$ are a fitting parameters. Note that the authors "do not have a good explanation of the factor $r^B$ and the numerical value of $B$" which  they used to fit the data.  It should also be noted that the fitting coefficients $A,$ $B,$ and $C$ are not universal. For a set of samples that differ only in the distance $L$ between the electrodes, they turn out to be different for samples that differ in the parameter $L$.


 It should be noted that almost all of the experimental work cited above used the shadow mask technique to fabricate SNS-based Josephson contacts. The resulting structures had the SN-N-NS bridge geometry shown schematically in Fig.~\ref{Sketch}. In this type of contact, the distance $L$ between the S-electrodes is not the geometric size of the weak bonding region. This region is delocalized and includes the areas of the N-film under the superconductors where current injection into the superconductor takes place. The presence of such delocalization was also qualitatively indicated by the experimental data. Indeed, substitution of the experimentally determined $E_{T}$ into the expression (\ref{Ic(EthS)}) gave an estimate of the size of the weak coupling region, which is larger than $L$. 
 
 In our opinion, it is the significant difference in the geometry of the SN and SN-NS interfaces that leads to an overestimation of the theoretical value of the Thouless energy calculated using rigid boundary conditions at the interface of the composite SN electrode with the N-film of the bridge.
The purpose of this article is to perform a detailed analysis of the proximity effect between the SN electrode and the N-film of SN-N-NS bridge structures and to derive an expression for the Thouless energy that takes into account both the finite transparency of their SN boundaries and the delocalization of the weak link region.

{Before going down this path, it is first necessary to define what we mean by the Thouless energy in (\ref{Ic(EthS)}) in the SN-N-NS structures under consideration. 
By its definition \cite{Edwards_1972,PhysRevLett.39.1167, THOULESS197493,PhysRevLett.76.1130}, $E_T$ characterizes the sensitivity of the energy state of the system to boundary conditions. That is, it correlates the existing characteristic scale of spatial changes in the system's parameters with its geometric dimensions. Unfortunately, at SN-N-NS junctions
the geometric size of the weak link region   
cannot be strictly determined. Moreover, a supercurrent across the junction is expressed by the sum over the Matsubara frequencies of the terms combined by the Green's function. Their characteristic scale of spatial variations in the N-film $\xi_\omega=(D/2\omega)^{1/2}$ depends on the Matsubara frequencies $\omega=\pi T(2n+1),$ $n=0; 1; 2...$
In the low temperature range $T \ll T_c$ this sum converges at $\omega \approx \pi T_c$. Thus, a wide range of ratios between $L$ and $\xi_\omega$ leads to a some uncertainty in the choice of a single characteristic scale of spatial changes in the structure. } 

{There is another interpretation of the Thouless energy. It says that two states which energies differ by more than the Thouless energy are correlated. Otherwise, they can be considered as independent single-particle states whose energies are not shared. In our particular case, we are not dealing with the energy of states, but with a set of Green's functions that determine the supercurrent. The difference in the nature of the spatial changes in these functions is determined by the Matsubara frequency $\omega=\pi T(2n+1)$. Therefore, the step at which the changes occur is equal to $2\pi T$. }

{In order to determine from a numerical solution of the Usadel equations or from experimental data the temperature $T_{Th}$ for the transition from a sharp increase to a smooth character of the current change with decreasing temperature, we can introduce the Thouless energy for our problem as $E_T=2\pi T_{Th}.$ The Thouless energy determined in this way gives us the temperature at which the structure transitions from a discrete to an integral representation of its properties, such as the order parameter and the superconducting current.}

{{It must be emphasized that the definition of $E_T=2\pi T_{Th}$ is absolutely equivalent to the standard definition of $E_{Th}$ in Eq. (\ref{Ic(EthS)}). 
}}
{Having determined the Thouless energy as described above, we can further use the relation (\ref{Ic(EthS)}) to estimate the characteristic scale of spatial variations in the structure as $L  = (D/2\pi T_{Th})^{1/2} =\xi$ and thus find that $L$ coincides with $\xi_\omega$ at the first Matsubara frequency $\omega=\pi T$ and $T=T_{Th}=E_{T}/2 \pi$, that is, with the maximum value among the scales $\xi_{\omega}$.}
{{In other words, the result of the above reasoning can be reformulated as follows: the Thouless energy is determined by expression (1). The value of the Thouless temperature $T_{Th}=E_T/2 \pi$ required for the processing of the experimental data follows from the equality of the geometrical size of the structure with the maximum value $\xi_{\omega}$ among the set of characteristic scales of the problem.}

{Thus, according to the established rule for the Thouless energy, $E_T$, in the considered SN-N-NS structures, we must proceed from the equality of the scale $\xi_\omega$ to the effective geometric size of the weak coupling region $L+\zeta_{\omega}$ 
{
\begin{equation}
\xi_\omega=L+2 \kappa \zeta_{\omega}
\label{Eth_equ}
\end{equation}}
at the first Matsubara frequency $\omega=E_T /2$. The $\zeta_{\omega}$ in (\ref{Eth_equ}) is the maximum value among characteristic scales of spatial changes in the N-layer underneath the S-films 
{and $\kappa$ is  a parameter which fixed the part of effective coherence length $\zeta_{\omega}$ by which the effective distance between the electrodes increases.}
Note that at $\zeta_{\omega}=0$ the expression (\ref{Eth_equ}) is transformed into the definition (\ref{Ic(EthS)}) of $E_{T}$, which was previously used to describe SNS sandwiches with rigid boundary conditions at the SN interfaces.  
}

{To determine $\zeta_{\omega}$, it is sufficient to solve the problem on the proximity effect between a semi-infinite N-film and an extended SN electrode.}

\section{Proximity effect between extended SN electrode and semi-infinite N-film}

We assume that the dirty limit conditions are satisfied in the superconductor and normal metal in the bridge, its SN boundaries have finite transparency,  the thickness of the N-film, $d$, is much less than $\xi_N=(D/2\pi T_c)^{1/2}$. 
{The suppression of superconductivity in the S-film due to the proximity effect with the N-layer is considered negligible and is not taken into account.}
Under these conditions, the proximity effect between the semi-infinite SN electrode $x\geq 0$ and the semi-infinite N-film $x\leq 0$ can be considered in the framework of the Usadel equations \cite{usadel1970}. In \cite{siegel2005density} it was shown that in the N-film under the S-electrode they can be written in the form:
\begin{equation}
\zeta_{\omega}^{2}\frac{\partial ^{2}}{\partial x^{2}}\theta-\sin (\theta
-\theta(\infty ))=0,~x\geq 0,  \label{TetNx}
\end{equation}
where
\begin{equation}
\theta(\infty )=\arctan \frac{\pi T_{c}\sin (\theta _{S})}{\omega
\gamma _{BM}+\pi T_{c}\cos (\theta _{S})},~  \label{TetNinf}
\end{equation}%

\begin{figure*}
	\centering
	\includegraphics[width=0.45\linewidth]{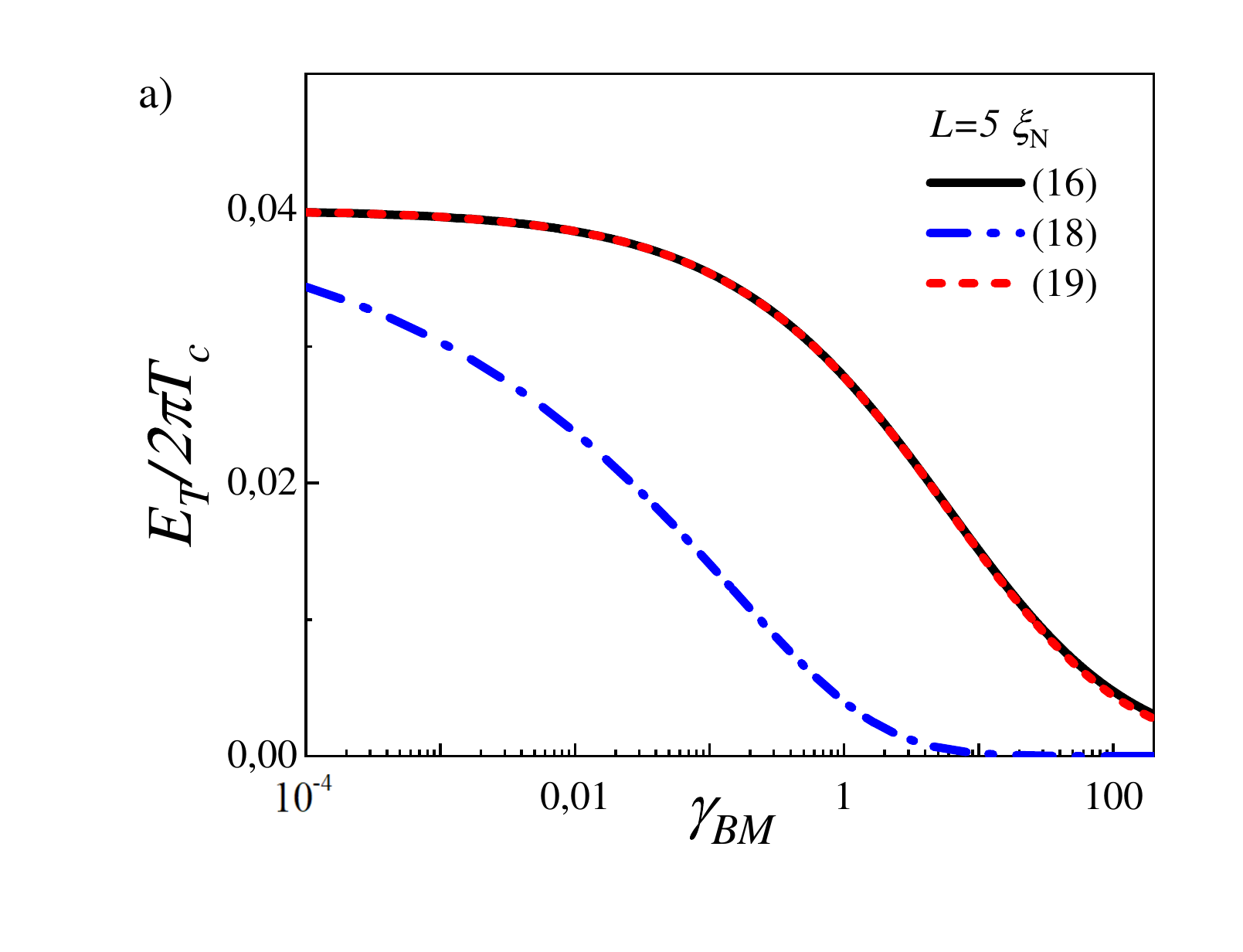}
	\includegraphics[width=0.45\linewidth]{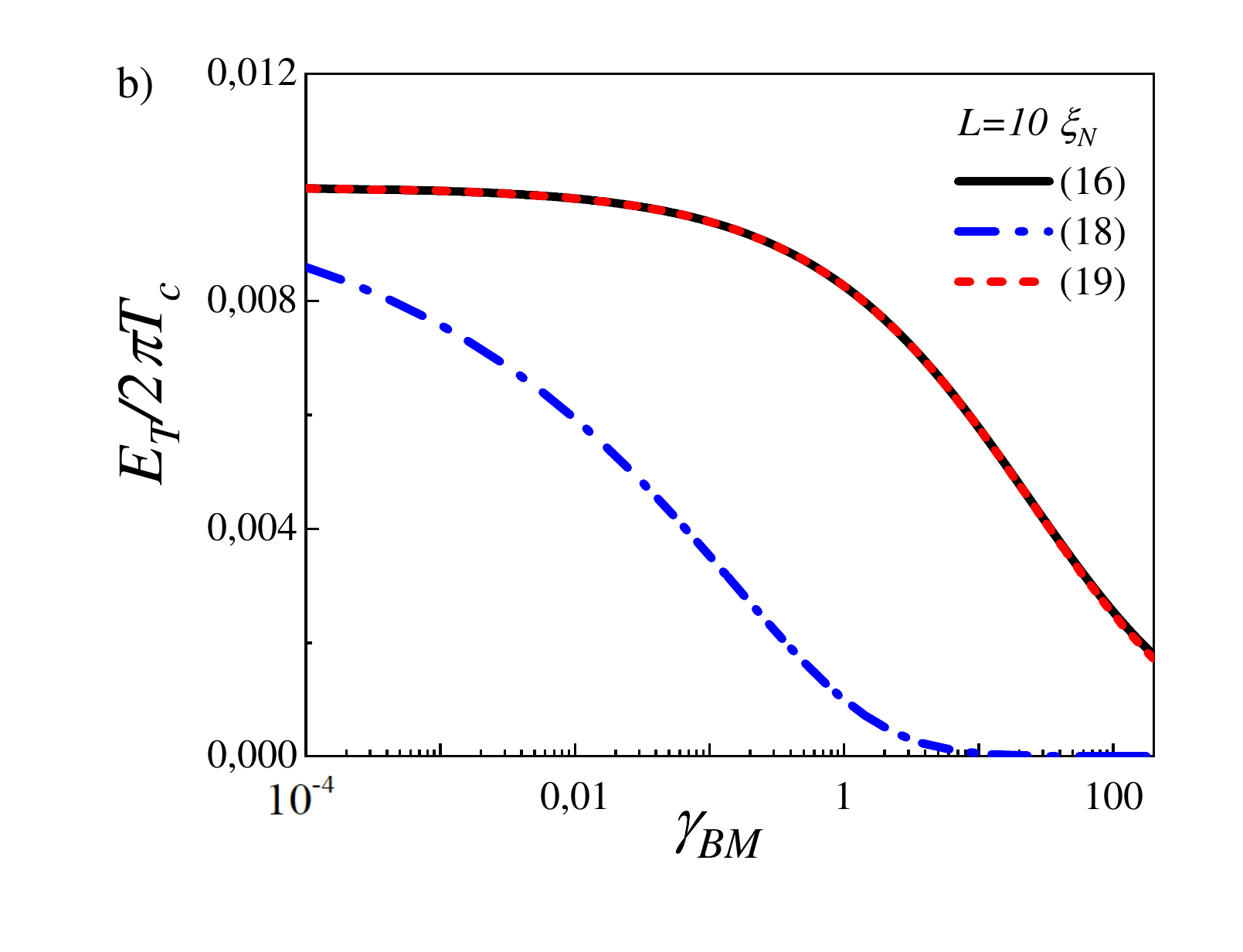} 
	\caption{ Thouless energy $E_{T}$ versus interface parameter $\gamma_{BM}$ in the SN-N-NS bridge with the distances between the electrodes $L=5\xi_N$ (a) and $L=10\xi_N$ (b), calculated from the Eq. (\ref{EqTh}) (solid line) and from the asymptotic expressions   (\ref{Lprom}) (dash-dotted line) and (\ref{ThGsm}) (dashed line), valid in the limit of large $\epsilon \gamma_{BM} \gg 1$ and small $\epsilon \gamma_{BM} \ll 1$ respectively.}
	\label{ETh_high}
\end{figure*}

\begin{equation}
\frac{\zeta_{\omega}}{\xi_{\omega}}= \sqrt{\frac{\omega \gamma
_{BM}}{\sqrt{\left( \omega \gamma _{BM}\right) ^{2}+2\pi T_{c}\omega \gamma _{BM}\cos
(\theta _{S})+\left( \pi T_{c}\right) ^{2}}}},
\label{dzitaDef}
\end{equation}%
\begin{equation}
 \cos (\theta _{S})=%
\frac{\omega }{\sqrt{\omega ^{2}+\Delta ^{2}}},~\sin (\theta _{S})=\frac{%
\Delta }{\sqrt{\omega ^{2}+\Delta ^{2}}},  \label{gammas}
\end{equation}%
and $\gamma_{BM}=\gamma_B d/\xi_N,$ $\gamma_B=R_B/\rho_N \xi_N,$ $R_B$ is the specific boundary resistance, $\Delta$ is  order parameter in the S layer. It has the BCS-like temperature dependence. 

In the N-film, the Usadel equation is
\begin{equation}
\xi_{\omega}^{2}\frac{\partial ^{2}}{\partial x^{2}}\theta-\sin (\theta
_{N})=0,~x\leq 0,  \label{TetFx}
\end{equation}%
The existence of the first integral of  equations (\ref{TetNx}) and (\ref{TetFx}) %
\begin{equation}
\zeta_{\omega} \frac{\partial }{\partial x}\theta=2\sin \left( \frac{\theta
_{N}(\infty )-\theta}{2}\right) ,~x\geq 0,
\end{equation}%

\begin{equation}
\xi_{\omega} \frac{\partial }{\partial x}\theta=2\sin \frac{\theta}{2}%
,~x\leq 0,
\end{equation}
permits to get their analytical solutions 
\begin{equation}
\theta=\theta(\infty )+4\arctan \left[ Q\exp \{-\frac{x}{\zeta_{\omega}%
}\}\right] ,~x\geq 0,  \label{SolTetN}
\end{equation}%
\begin{equation*}
Q=\tan \frac{\theta(+0)-\theta(\infty )}{4}
\end{equation*}%
\begin{equation}
\theta =4\arctan \left[ \left( \tan \frac{\theta (-0)}{4}\right) \exp
\left\{ \frac{x}{\xi _{\omega }}\right\} \right] ,~x\leq 0,  
\label{SolTetF}
\end{equation}%
The integration constants $\theta(\pm0)$  in (\ref%
{SolTetN}), (\ref{SolTetF}) 
\begin{equation}
\theta (\pm 0)=2\arctan \frac{\sin \frac{\theta(\infty )}{2}}{\cos \frac{%
\theta(\infty )}{2}+g},\quad g=\frac{\xi_{\omega}}{\zeta_{\omega}}.
\label{Tetpm}
\end{equation}%
have been determined from the boundary
conditions %
\begin{equation}
\frac{\partial }{\partial x}\theta(+0)=\frac{\partial }{\partial x}%
\theta(-0)  \ \text{and} \ \theta(-0)=\theta(+0).  \label{BCx=0gB}
\end{equation}%
at SN-N interface $(x=0)$.

\section{Thouless energy}

{Substitution of (\ref{dzitaDef}) into (\ref{Eth_equ}) at the first Matsubara frequency $\omega= E_T /2$ results in
{
\begin{equation}
\frac{L}{\xi _{N}}+ 2 \kappa \sqrt{\frac{%
\gamma _{BM}}{\sqrt{\epsilon ^{2}\left( \gamma %
_{BM}^{2}+2\gamma ^{\ast }\gamma _{BM}\right) +1}}}=%
\sqrt{\frac{1}{\epsilon }},
\label{EqTh}
\end{equation}}
where $\epsilon=E_{T}/ 2\pi T_c$ 
and we replace $\Delta$ by its value $\pi T_c/\gamma ^{\ast },$ at $T \ll T_c$; $\gamma^{\ast }\approx 1.781$ is Euler's constant.}

{In the limit $\epsilon \gamma_{BM} \gg 1$ the equation (\ref{EqTh}) transforms to: }
{
{\begin{equation}
\epsilon =\frac{\xi _{N}^{2}}{L^{2}}\left( 1- 2 \kappa \sqrt{%
\frac{\gamma _{BM}}{\sqrt{\left( \gamma _{BM}^{2}+2%
\gamma ^{\ast }\gamma _{BM}\right) }}}\right) ^{2}.
\label{Lprom}
\end{equation}}}
{Note that at $\gamma _{BM}\rightarrow \infty ,$ 
the SN boundaries become completely opaque for quasiparticles located in the N-region. The current located in the N-film of the SNS contact cannot flow into the S-electrodes. If the length of the electrode SN boundary 
significantly exceeds $\zeta_{\omega}$, the length of the current localization region in its N-part can be considered as infinite. 
 Therefore in the full agreement with (\ref{Ic(EthS)}) the parameter $\epsilon$ should tends to zero and $\gamma _{BM}\rightarrow \infty$.
From this requirement we get $\kappa =1/2$ and
\begin{equation}
\epsilon =\frac{\xi _{N}^{2}}{L^{2}}\left( 1-\sqrt{\frac{\gamma _{BM}}{\sqrt{\gamma _{BM}^{2}+%
2\gamma^{\ast } \gamma _{BM}}}}\right) ^{2}.
\label{ThLgamma}
\end{equation}
}
{{Suppose further that parameter $\kappa$ is independent on $\gamma_{BM}$ 
in the opposite limit then $\epsilon \gamma_{BM} \ll 1$ we get
\begin{equation}
\epsilon=\frac{\xi _{N}^{2}}{(L+\xi _{N}\sqrt{\gamma _{BM}})^2}.
\label{ThGsm}
\end{equation}}}

Figures \ref{ETh_high}a,b show the dependence of the Thouless energy $E_{T}$ on the suppression parameter $\gamma_{BM}$, calculated with the equations (\ref{EqTh}), (\ref{Lprom}), (\ref{ThGsm}), for the distances between the electrodes SN-N-NS structures equal to $L=5\xi_N$ and $L=10\xi_N$. The calculation shows that the approximation formula (\ref{ThGsm}) agrees quite well with the exact dependence $E_{T} (\gamma_{BM})$ calculated by the formula (\ref{EqTh}) in the considered range of parameters for long bridges. A noticeable discrepancy between the curves calculated by the formulas (\ref{EqTh}) and (\ref{ThGsm}) appears only in the $\gamma_{BM}\gg 100$ limit, which is not actually realized in the experiments of SN-N-NS structures.

At the same time, the expression (\ref{Lprom}) fits the exact dependency $E_{T} (\gamma_{BM})$ rather badly. The reason is that the condition $\epsilon \gamma_{BM} \gg 1$ is not fulfilled even in the region of large $\gamma_{BM}$. This is due to the fact that $\epsilon$ decreases much faster than $\gamma_{BM}^{-1}$. The difference between the exact solution (\ref{EqTh}) and that resulting from asymptotic expression (\ref{Lprom})  becomes more obvious as $L/\xi_N$ increases.

{It is easy to see that when $L \gg \xi_N \sqrt{\gamma_{BM}}$ the expression (\ref{EqTh}) goes to (\ref{Ic(EthS)}). Physically, this limit is equivalent to the fulfillment of rigid boundary conditions at the N boundaries of the bridging film with its NS electrodes. In this limit, as in the SNS sandwiches, investigated in \cite{dubos2001josephson}, the rigid boundary conditions do not impose any additional characteristic lengths characterizing the 
spatial variations of parameters in the N film of SNS or SN-N-NS structures. The weak coupling region turns out to be a “closed” system, 
characterized only by its intrinsic parameters, for example, by the diffusion coefficient $D$. The Thouless energy (\ref{Ic(EthS)}) was introduced  precisely for such “closed” systems.
}

{Going beyond the rigid boundary conditions, for example, taking into account the finite transparency of SN boundaries in SNS sandwiches or delocalization of the weak coupling region occurring in the studied SN-N-NS structures, violates the closure condition. Additional scales appear in the problems. They are determined by the suppression parameter $\gamma_B$ in SNS sandwiches or $\gamma_{BM}$ in SN-N-NS structures. Thus, in SN-N-NS bridges in accordance with the formula (\ref{EqTh}) at $L \ll \xi_N \sqrt{\gamma_{BM}}$ the diffusion coefficient ceases to characterize the system and the Thouless energy $E_T = \pi T_c/ \gamma_{BM}$ in the first approximation is determined only  by the parameter  $\gamma_{BM}$, i.e., by the properties of the SN boundaries.
}

{The proposed definition (\ref{EqTh})  of $E_T$ is a compromise solution,
taking into account both the internal parameters of the N-metal in the weak coupling region and the peculiarities of the method of inducing
superconducting correlations in it.
}

\begin{figure*}
	\centering
	\includegraphics[width=0.45\linewidth]{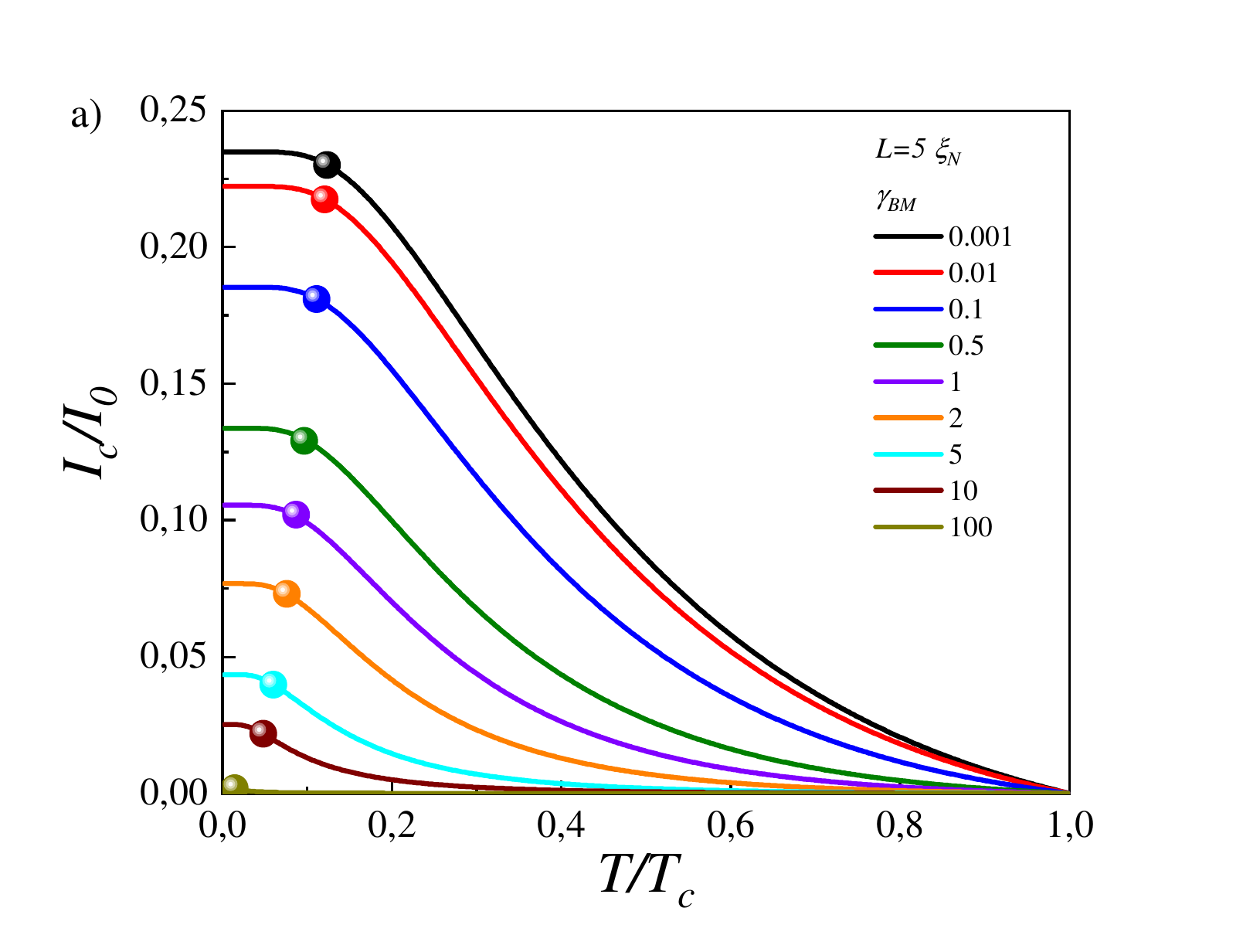}
    \includegraphics[width=0.45\linewidth]{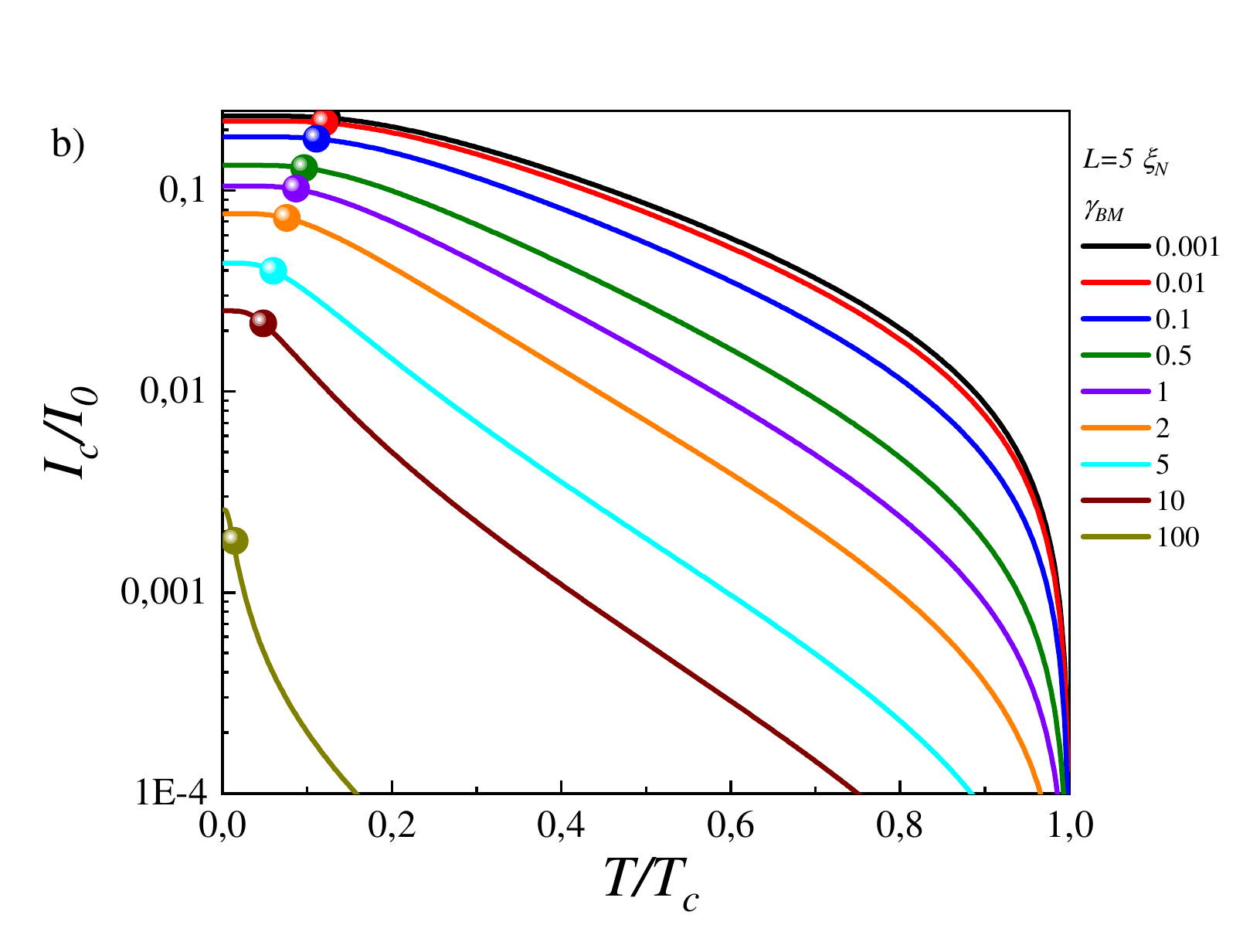}
    \vfill
    \includegraphics[width=0.45\linewidth]{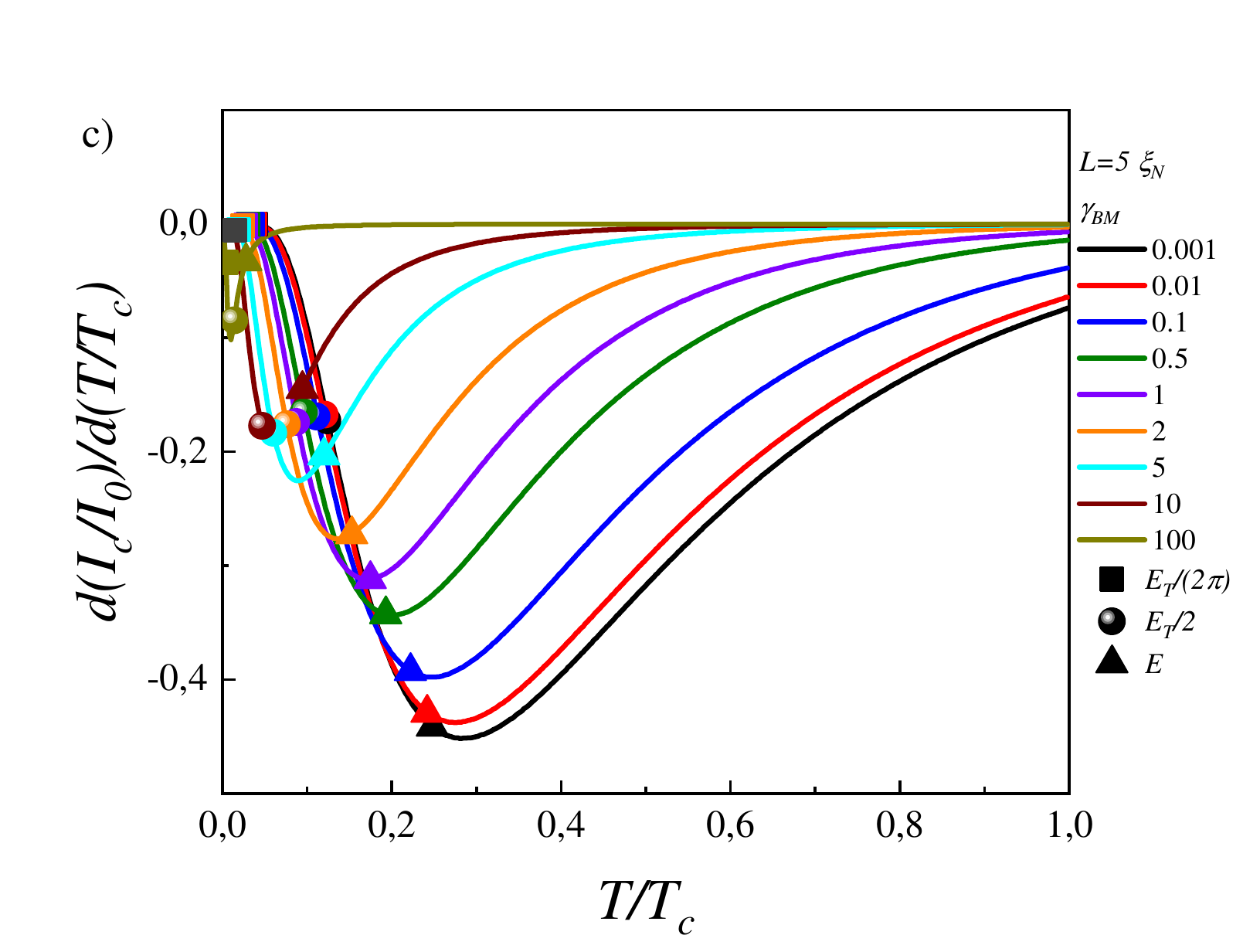}
    \includegraphics[width=0.45\linewidth]{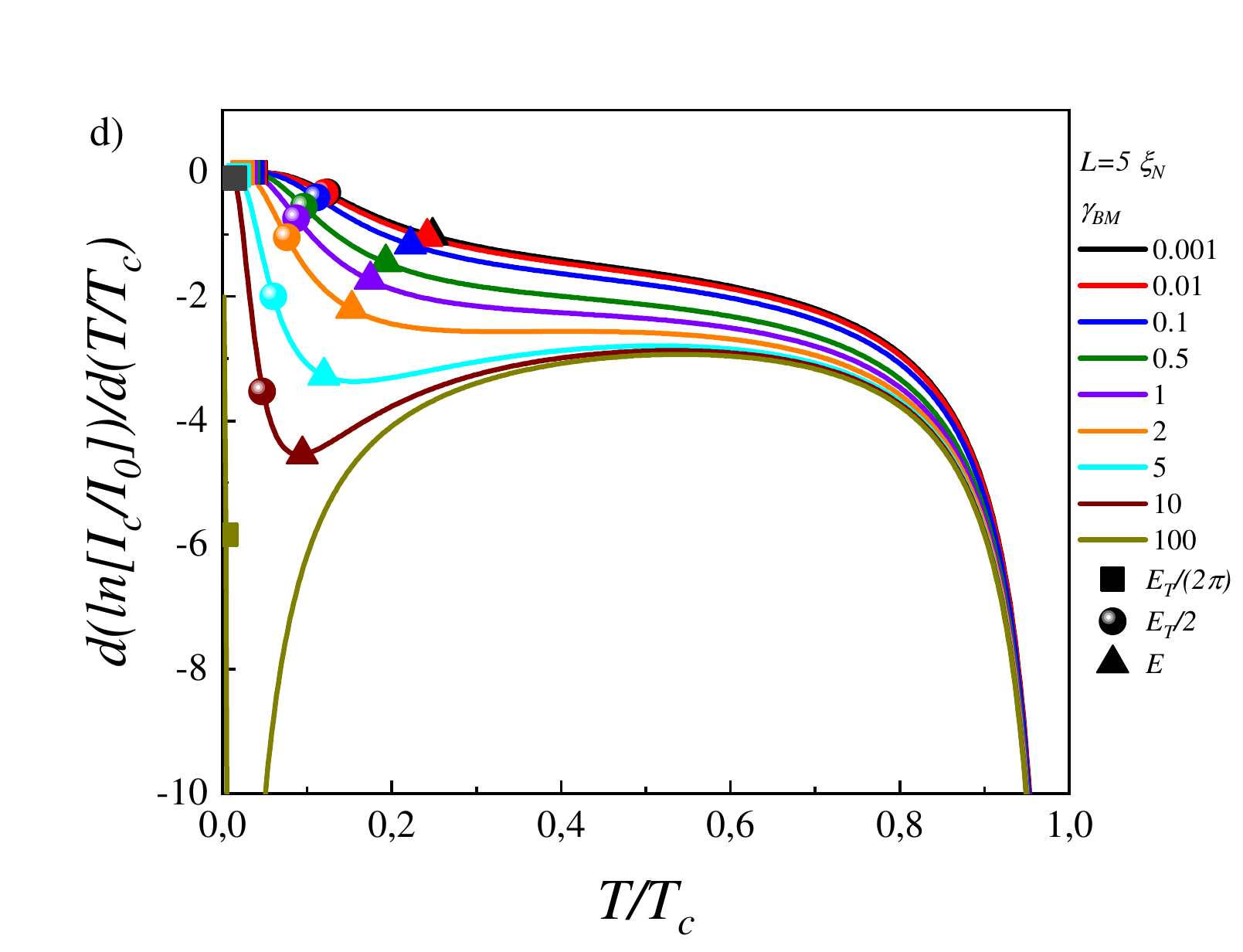}
    \vfill
    \includegraphics[width=0.45\linewidth]{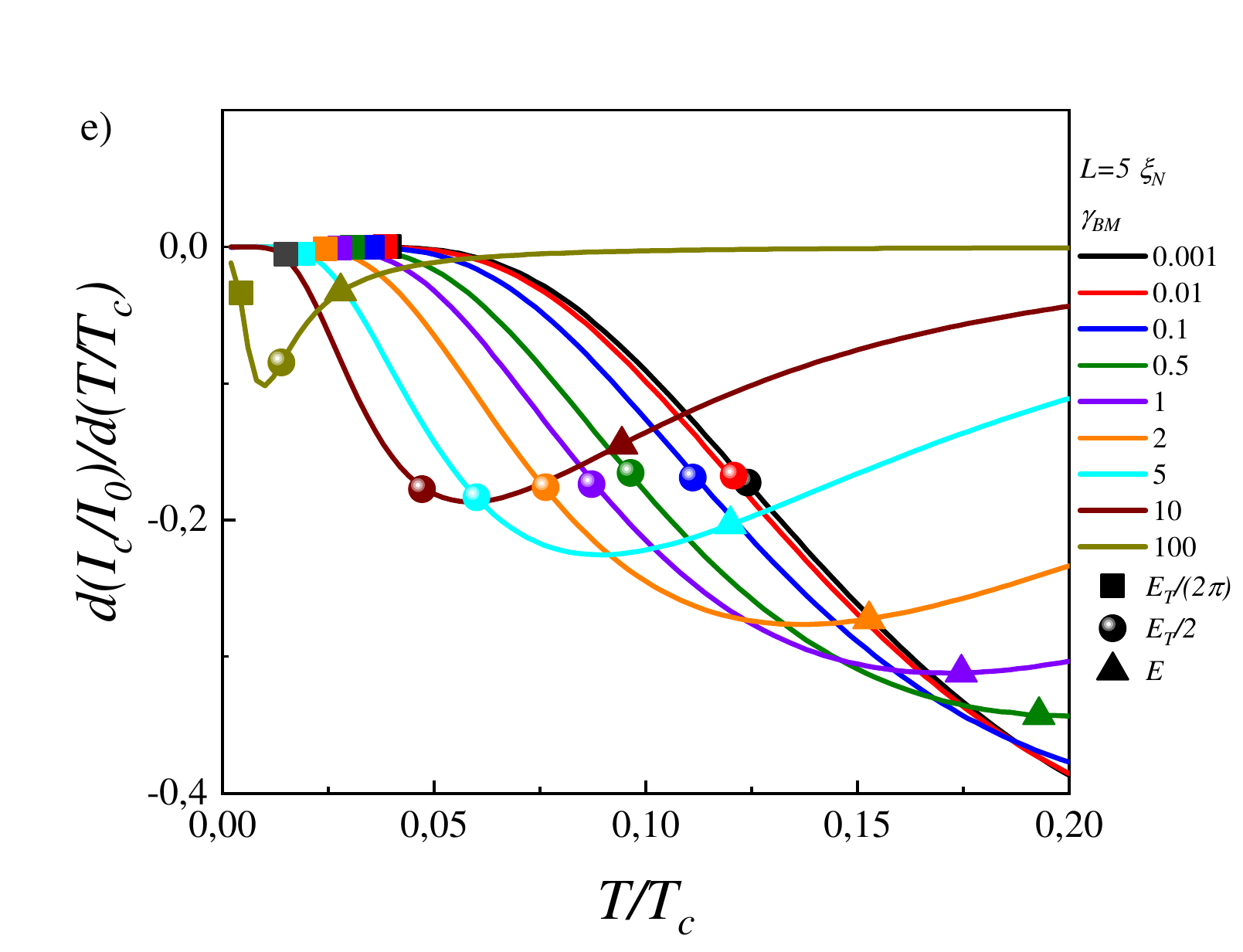}
    \includegraphics[width=0.45\linewidth]{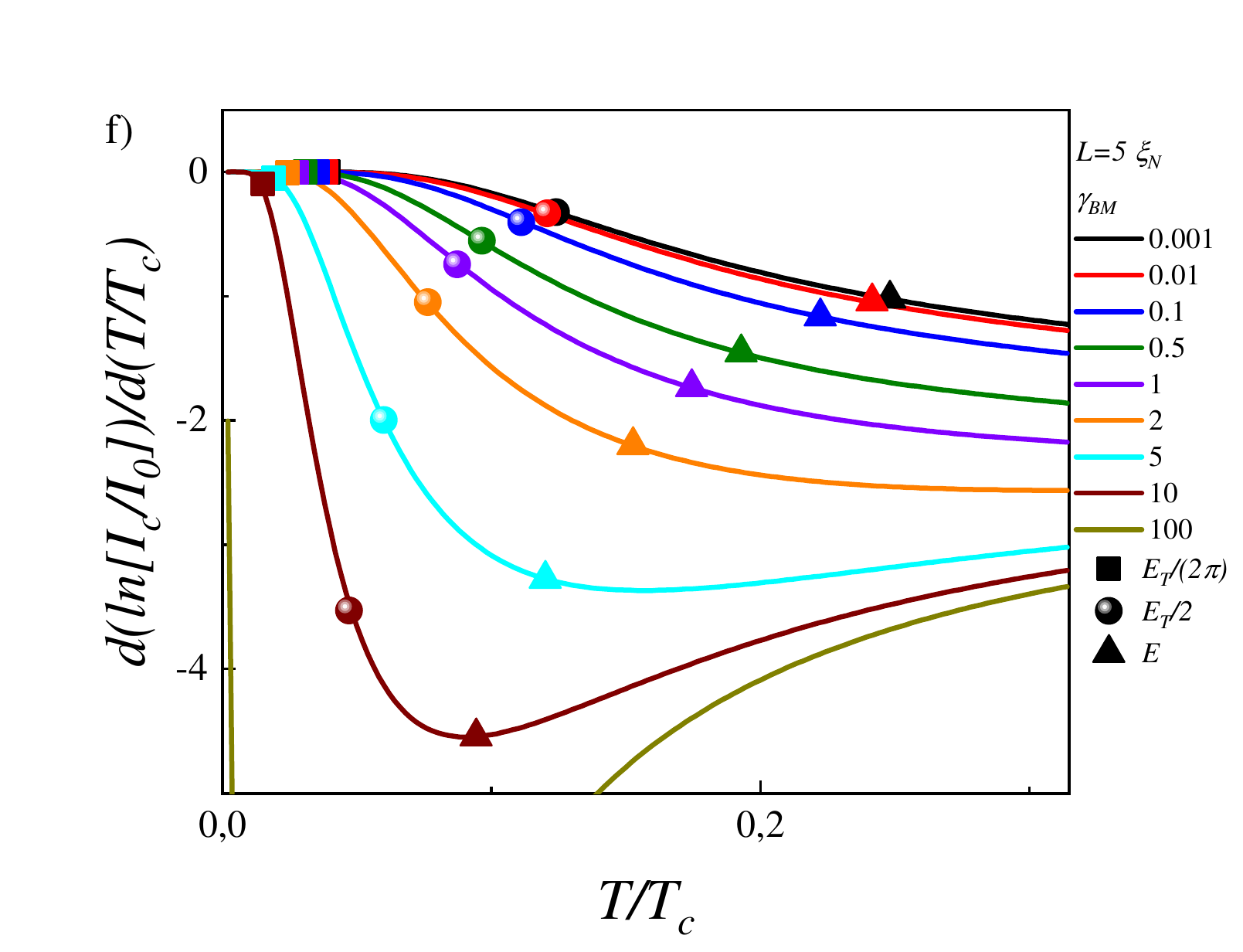}
	\caption{{Temperature dependence of the critical current density $I_c$ of the SN-N-NS bridge with $L=5\xi$ calculated numerically for set of different interface parameters $\gamma_{BM}$ in linear (a) and logarithmic (b) scale. Panels (c) and (d) reveal the temperature dependence of the first derivative of the critical current density $\frac{dI_c/I_0}{dT/T_C}$ at the same parameters in linear and logarithmic scale respectively. The {squares, points and triangles} on each of the calculated curves show the temperature values corresponding to {$E_T/2\pi$}, {$E_T/2$} and {$E_T$}, respectively. The values of {$E_T$} were calculated using the expression (\ref{ThGsm}) for each combination of $\gamma_{BM}$ and $\xi_N$ parameters. Panels (e) and (f) is enlarged version of (c) and (d) in the low temperature region of parameters.}}
	  \label{JCT}
\end{figure*}

{\section{Calculation of a supercurrent across SN-N-NS junction. }}
{The dependence of the superconducting current density, $I$ on the phase difference of the order parameters of the superconducting electrodes $\varphi$ in the SN-N-NS junctions has been previously calculated by numerical methods in \cite{soloviev2021miniaturization} for arbitrary values of $\gamma_{BM}$ and the ratio $L/\xi_N$.
Just as in the approach we used earlier in Section II, the suppression of superconductivity in the S-film was not taken into account in [9], and for computational convenience the Usadel equations were written in the $\Phi$ representation (the origin of the $0x$ axis is placed in the center of the structure). 
\begin{equation}
\xi_{eff}^{2}\frac{\partial }{\partial x}\left( G^{2}\frac{\partial \Phi }{
\partial x}\right) -\Phi =-\delta,\  x \geq L/2
\label{EqInN}
\end{equation}%
\begin{equation}
\frac{\partial }{\partial x}\left( G^{2}\frac{\partial \Phi }{\partial x}
\right) -\Omega G\Phi =0,\ 0\leq x\leq L/2  \label{EqInN1}
\end{equation}%
\begin{equation}
\xi _{eff}^{2}=\frac{\gamma _{BM}}{G\left( G_{s}+\gamma _{BM}\Omega \right) }%
,~\delta =\frac{G_{s}\Delta \exp{i\varphi /2}}{\left( G_{s}+\gamma _{BM}\Omega \right) }.
\label{ksieff}
\end{equation}%
\begin{equation}
\frac{I}{I_0}=\frac{2\pi T}{T_c}\sum_{\Omega \geq 0}^{\infty }\frac{G^{2}}{\Omega ^{2}}
\left( \mathrm{Im}\Phi \frac{\partial \mathrm{Re}\Phi }{\partial x}-\mathrm{
Re}\Phi \frac{\partial \mathrm{Im}\Phi }{\partial x}\right),
\label{currentN}
\end{equation}
Here, $\Phi $ and $G=\Omega /(\Omega ^{2}+\Phi \Phi ^{\ast })^{1/2}$ are Usadel Green's functions, the  Matsubara frequencies
$\Omega =(2n+1)T/T_{c}$ are normalized
on $\pi T_{c}$,  the coordinate $x$ is normalized on $\xi _{N}$ and $I_0= T_{c}/e\xi _{n}\rho _{N}$, $\rho_N$ is resistivity of the N-film.
The modulus of the order parameter in the S-electrode $\Delta $ has the BCS-like temperature dependence and is normalized
on $\pi T_{c}$. 
 }

{ Equations (\ref{EqInN}), (\ref{EqInN1}) should be supplemented by boundary conditions. At $x=0$ they follow from the symmetry of the considered problem
\begin{equation}
\frac{\partial \mathrm{Re}\Phi }{\partial x} =0,\ \mathrm{Im}=0,
\label{BC0}
\end{equation}
At a large distance from the N-NS boundary, the functions $\Phi$ converge to a solution that does not depend on x
\begin{equation}
\Phi=\delta,
\label{BCinf}
\end{equation}
}

Figures \ref{JCT}a,b show the $I_c(T)$ dependencies computed for $L=5\xi_N$ and for the set of different interface parameters $\gamma_{BM}$ in the range from $0.01$ to $100$. For convenience, they are presented as linear and logarithmic scales. The points on the curves mark the position on the {$T/T_c = E_{T}/2$} axis calculated for the given combination of $L=5\xi_N$ and $\gamma_{BM}$ using Eq. (\ref{ThGsm}). 
The figures clearly show that at these temperatures the slope of the dependencies $I_c(T)$ in Fig. \ref{JCT} changes, i.e. there is a transition from a smooth to a sharp temperature drop of $I_c(T)$ with increasing temperature.

{In addition to the temperature $T=E_T /2$, two other characteristic temperature values can be noted. They are $T=T_{Th}=E_T/2 \pi$  marked with squares and $T=E_T $ marked with triangles on the temperature dependencies of the first derivative of the critical current with respect to temperature (see Fig.\ref{JCT}c-f). Figures \ref{JCT}d,f clearly show that for $L=5\xi_N$ and $\gamma_{BM}=0.01$, the Thouless temperature is $T_{Th}\approx 0.04 T_c$. The increase of the parameter $\gamma_{BM}$ is accompanied by a shift of the values of this temperature towards $T=0$, so that at $\gamma_{BM}=2$, the value of $T_{Th}$ turns out to be equal to $\approx 0.025 T_c$. At $T=T_{Th}$ the critical current reaches saturation. So the first derivative at $T \lesssim T_{Th}$ is $dI_c/dT=0$. From the presented in Fig.\ref{JCT}d,f  dependencies it also follows that there are extra two points of temperature. {At $T \approx E_{T}$ and $\gamma_{BM} < 5$ the local minimum of the derivative is reached. At $T \approx E_{T}/2$ an inflection point of $dI_c/dT$ appears  for $\gamma_{BM} < 5$. Qualitatively, this point corresponds to the transition area between saturation region and growing part of $I_C(T)$ dependence.  }
Reference points at $T=E_{T}/2$ and $T=E_{T}$ can be useful when values of $T=T_{Th}$ are difficult to achieve due to limitations on the temperature range allowed for measurements.  
}
{In the limit $\gamma_{BM} \gg 5 $ the effective impact of boundaries increases and the positions of the considered points $T \approx E_{T}/2$ and  $T \approx E_{T}$ are shifted}

{Note that in the limit of small critical current, the boundary value problem (\ref{EqInN})-(\ref{BCinf}) may also have an analytical solution. This limit is realized in the case when it is possible at any coordinate $x$ to neglect the suppression of anomalous Green's functions induced in the N-film by the current flowing through it. }

\section{Critical current of the long SN-N-NS bridge}

In the $L\gg \xi$ limit, the superconducting state near the center of the bridge, $\Phi(x),$ can be described by the superposition of anomalous Green's functions (\ref{SolTetF}), (\ref{Tetpm}) penetrating from the superconducting banks into the bridge \cite{likharev1975steady,zaikin1981theory,lukichev1982}:
{\begin{equation}
\frac{\Phi}{\omega}=\tan \left[ \theta(-x-\frac{L}{2}) \right ] e^{-i\frac{\varphi}{2}} 
+\tan\left[ \theta(x-\frac{L}{2}) \right] e^{i\frac{\varphi}{2}},
\label{ssuperpos}
\end{equation}%
where $\theta(x)$ is the solutions of the proximity problem (\ref{SolTetN})-(\ref{Tetpm}). By substituting this solution of the Usadel equations into the expression for the supercurrent (\ref{currentN}), we arrived at a sinusoidal dependence of the supercurrent on $\varphi$ with a critical current density equals to}    
\begin{equation}
\frac{I_{c}}{I_{0}}=\frac{128\pi T}{T_{c}}\sum_{\omega }\tan
^{2}\left( \frac{1}{2}\arctan \frac{\sin \frac{\theta(\infty )}{2}}{%
\cos \frac{\theta(\infty )}{2}+g}\right) e^{-\frac{L}{\xi_{\omega}}}. 
\label{JC}
\end{equation}

In the considered approximation 
{($L \gg \xi$)}
the critical current value is determined by the first term in the sum in (\ref{JC}):
\begin{equation}
\frac{I_{c}}{I_{0}}=\frac{128\pi T}{T_{c}}\tan
^{2}\left( \frac{1}{2}\arctan \frac{\sin \frac{\theta(\infty )}{2}}{%
\cos \frac{\theta(\infty )}{2}+g}\right) \exp \{-\frac{L}{\xi }%
\}, 
\label{JC11}
\end{equation}%
{where 
$\theta(\infty )$ is determined by Eq. (\ref{TetNinf}) with $n=0$, that is $\omega=\pi T$.} 
The expression (\ref{JC11}) is an analog of the formula (\ref{Ic(EthL)}) derived in \cite{zaikin1981theory} for SNS structures with rigid boundary conditions at the SN boundaries. Unlike (\ref{Ic(EthL)}), the expression (\ref{JC11}) takes into account both the finite transparency of the SN boundaries and the delocalization of the weak coupling region. 

{Note, that the expression (\ref{JC}) is also valid in limit of large $\gamma_{BM} \gg 1$, if the distance $L$ between the SN electrode is not too small, so that the sum in (\ref{JC}) converges at
\begin{equation}
\omega \lesssim \frac{\pi T_c}{\gamma_{BM}} \max \left[ 1, \frac{L}{\xi_N} \right].    
\label{condLGBM}
\end{equation}
}

{\section{Comparison with experimental data}
}

{From the theoretical results given above it follows that it is possible to determine $T_{Th}$ on the basis of the available (calculated or experimentally obtained) dependence $I_c(T)$ and, using the formula (\ref{ThGsm}), to find such an important parameter for practical applications as the effective geometric size of the SN-N-NS Josephson contacts.
\begin{equation}
L_{ef} = L+\xi_N \sqrt{\gamma_{BM}}. 
\label{Leff}
\end{equation}%
}

{Note that Eq. (\ref{Leff}) is the direct consequence of Eq. (\ref{ThGsm}) and is valid up to $\gamma_{BM}\approx 100 $, as can be seen from the comparison of the exact (\ref{EqTh}) and asymptotic (\ref{ThGsm}) solutions for $E_T$ in Fig. \ref{ETh_high}.} 

{Our experimental study of a supercurrent transport across Nb/Cu-Cu-Cu/Nb \cite{skryabina2017josephson} and Nb/Au-Au-Au/Nb \cite{skryabina2021environment, sotnichuk2022long} bridges had shown that the shape of the exponential dependencies of $I_c(T)$ in a wide temperature range was close to that shown in Fig.\ref{JCT}.
At $\gamma_{BM} \lesssim 5$ and large values of $L\gtrsim 4\xi_N$ with decreasing $T$, the transition from a sharp rise of $I_c(T)$ to a smoother saturation at $T \ll T_c$ occurs at $T\approx T_c (\xi_N /L)^2$.  At $\gamma_{BM} \gtrsim 5$ and small values of $L\lesssim 4 \xi_N$ the change in $I_c(T)$ takes place at $T\approx T_c / \gamma_{BM}$.
}

{In Figure \ref{FigExp}a we show the latest experimental data for the Nb/Au-Au-Au/Nb Josephson bridge with a diffusive Au stripe as the weak link. This structure was fabricated by magnetron sputtering with lift-off lithography.  The details of the fabrication process can be found in \cite{skryabina2021environment, sotnichuk2022long, ElistratovaFuture}. The approximate distance between the SN electrodes can be estimated to be about $L \approx 160$ nm from the scanning electron microscope image (Fig. \ref{Sketch}a) of this bridge. The thicknesses of the niobium and gold layers are about $70$ nm and $32$ nm respectively.}

\begin{figure}[t]
	\centering
    \includegraphics[width=0.95\linewidth]{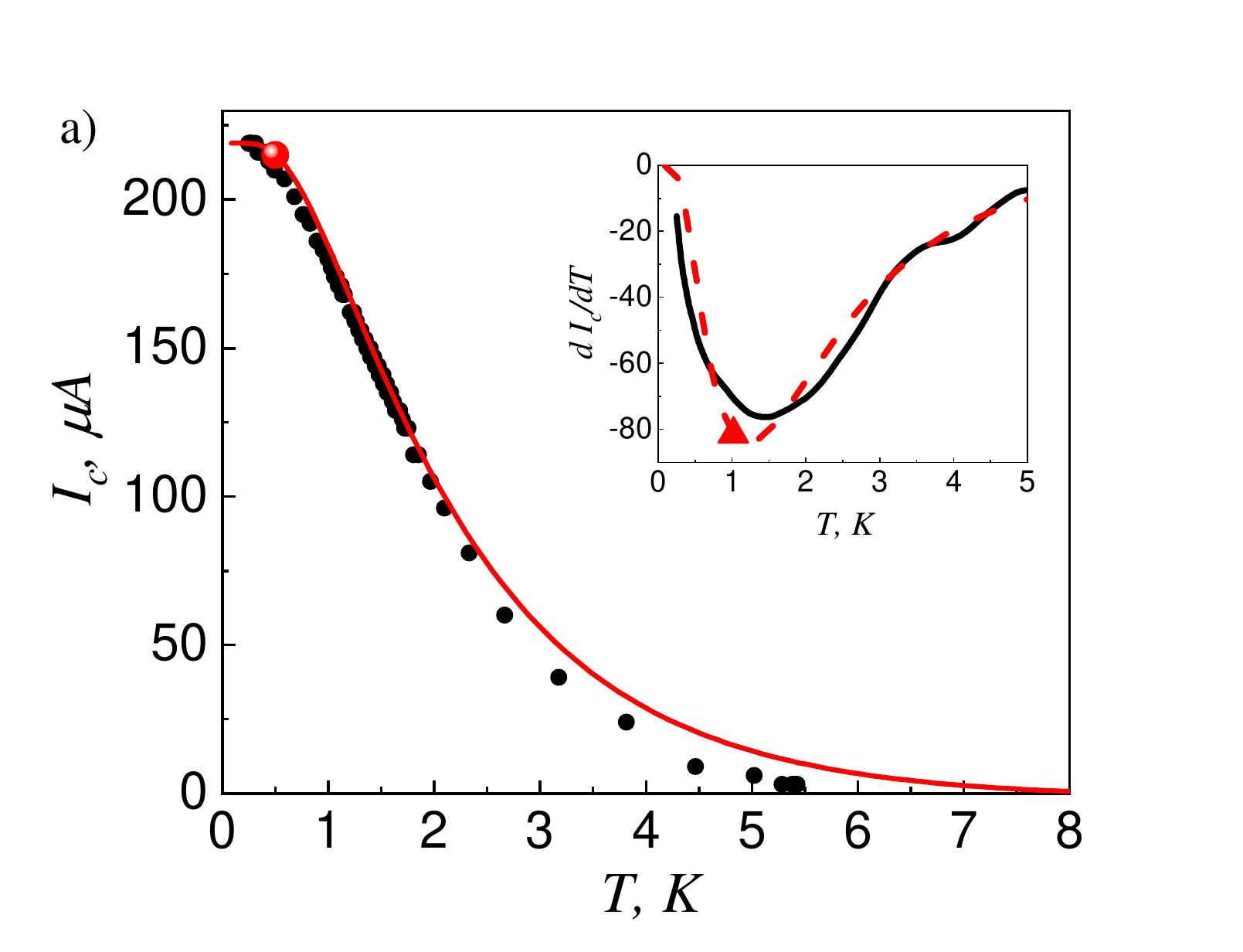}
    \includegraphics[width=0.95\linewidth]{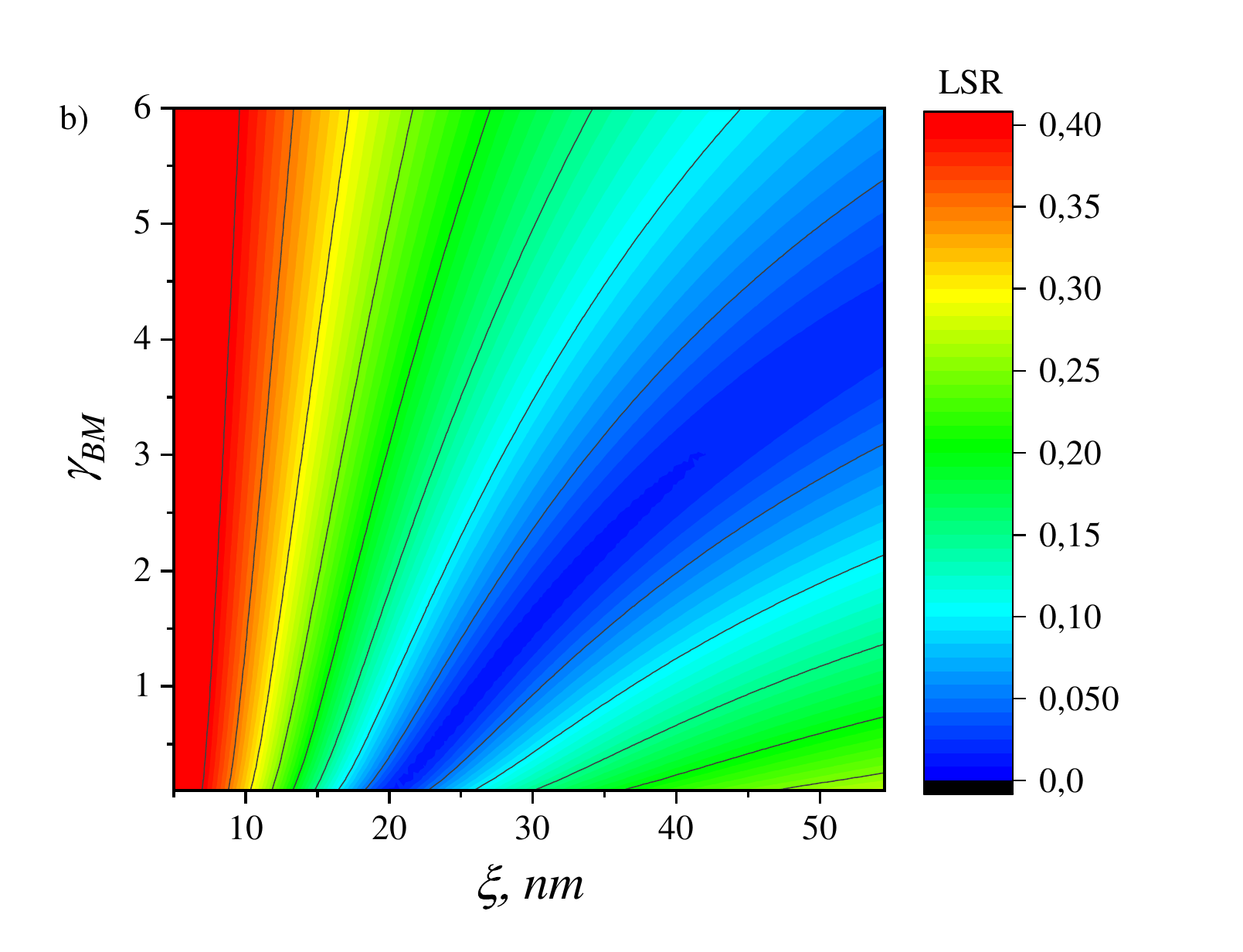}
	\caption{{a) Temperature dependence of the critical current $I_c(T)$ in the NbAu-Au-AuNb bridge with the length of the weak link $L=160$ nm (black circles) and numerical fits in the framework of the model (\ref{EqInN})-(\ref{BCinf}) calculated at $\xi_N=23$ nm and $\gamma_{BM}=0.12$ (red line). The red dot marks the temperatures corresponding to the {halfed} Thouless energy $E_T/2$. The inset shows the derivative $dI_c(T)/dT$ of the interpolated experimental (black solid line) and model (red dashed line) $I_c(T)$ dependencies. b) The least-squares residual between the experimental data and the model fits in the $(\xi,\gamma_{BM})$ plane of the parameters. }}
	\label{FigExp}
\end{figure}


{To compare the experimental data presented in Fig. \ref{FigExp}a with the results of the theoretical calculations, we construct a map of the dependence of the standard deviation of the experimental data on the calculated values. The least-squares calculation of the standard deviation of the theoretical values of the critical current from the experimentally obtained points is shown in Fig. \ref{FigExp}b. The result of this procedure is the detection of a valley of values in the region of the parameters $\gamma_{BM}$ and $\xi$ where the experimental points agree quite well with the theoretical calculation.  }

{It can be seen that the presence of two free parameters $\gamma_{BM}$ and $\xi$ does not lead to the uniqueness of the solution of the fitting data. For different sets of $\gamma_{BM}$ and $\xi$ the formula (\ref{ThGsm}) gives different values of $E_T$. 
}

{It should be noted that in the particular case we are considering, the expression (\ref{ThGsm}) allows us to specify the range of parameters in which it is worthwhile to search for a solution of the system at sufficiently large effective bridge lengths $L \gtrsim 5 \xi$ and interfaces transparent enough $\gamma_{BM} < 1$. As an example, in Fig. 4a we have shown one of the options of the approximation of the experimental dependence $I_c(T)$ with the solution of the Usadel equations obtained at parameters $\xi_N=23$ nm ($L/ \xi_N \approx 7)$ and $\gamma_{BM}=0.12$ (red line).
A large red dot on this graph marks the temperature 
{$T\approx 0.5K$}, which corresponds to the value $E_T/2$ (see (\ref{ThGsm})).
The inset of Fig. 4a shows the derivative of the theoretical $dI_c/dT$ dependence (red dashed line) and the interpolation curve obtained for the experimental points. The red triangle also marks the point where the temperature {$T\approx 1 $K} coincides with the value corresponding to $E_T$. It is located near the minimum of the derivative $dI_c/dT$.   
}

{It should be emphasized that according to (\ref{Leff}) for $L/\xi_N \approx 7$ and $\gamma_{BM}=0.12$ the effective size of the structure $L_{ef}\approx 7.3 \xi_N$ is actually determined by the distance between the electrodes $L$.
In this case, the features of the form of dependence $I_c(T)$ are due to the transition from exponential to power laws of growth $I_c$ with decreasing temperature. As can be seen in Fig.1, the position of these features correlates quite well with $T_{Th}$.
}

{For bridges with short and medium distances between the electrodes $L\lesssim 3\xi$, the $I_c(T)$ dependencies are smoother and the curves bending $dI_c(T)/dT$ at $T=E_{T}/2$ are not as clearly defined as for large values of $L$. Also in these case the interface parameter $\gamma_{BM}$ is going to raise over the length, changing the shape of the $I_c(T)$ dependence.
}



{Thus, we have shown that the expression (\ref{ThGsm}) can be used to determine the range of parameters in which the optimal approximation of the experimental results should be sought.}

{For a more unambiguous extraction of the structure parameters from the dependence $I_C(T)$, it is necessary to approximate several samples at once with different lengths between the electrodes $L$, as done in \cite{skryabina2017josephson, sotnichuk2022long} and using additional data obtained from the determination of the normal contact resistance. This approach allows one to significantly limit the set of parameters at which the measured dependencies are best described by the microscopic model. 
However, it requires a complete construction of the parameter map, which is a rather resource-intensive procedure.
}

{It can be seen that the theoretical curve fits the experimental data well at $T \lesssim 4$K. The observed discrepancy between the theoretical predictions and the experimental data at higher temperatures ($T \gtrsim 4$K) may be due to a number of reasons. 
}

{Technological reasons include a possible difference in the critical temperature of the region of the S layer adjacent to the N-film, $T_{cl}$ compared to the measured $T_c$ of the whole S electrode, as well as possible $T_{cl}(x)$ dependence. The next factor may be the inhomogeneity of the transparency of the NS boundaries, caused both by mechanical stresses at the NS interfaces, their roughness, and by the lift-off lithography process. 
All these factors lead to the appearance of island-type superconductivity in the N-layer, which is more pronounced the closer $T$ is to $T_c$, i.e. to the appearance of the dependence $\delta(x)$ in (\ref{ksieff}), which was not considered in the developed theory.
}

{It should be noted that such a successful coincidence of Thouless energy values found from the numerically calculated dependencies $I_c(T)$ and based on expression (2) does not mean that the values of $E_T$ determined from the shape of the experimental curves $I_c(T)$ will lead to value of the effective geometric size of the contact, $L_{ef},$ strongly determined by Eq. (\ref{Leff}).
A deviation of experimentally determined value of $L_{ef}$ and that follows from (\ref{Leff}) may be due to the fact that a number of constraints that determine the applicability of the one-dimensional SN-N-NS contact model we have used for our calculations were not met in an experiment.}

{Such restrictions include:}

{1. The smallness of the N-layer thickness compared to decay length $\xi_N$.  When using low resistivity normal metals (Al, Au, Ag), the proximity of $d$ to $\xi_N$ can lead to violation of the condition of absence of superconductivity suppression in the S-layer near the SN boundary \cite{orlikovskii1986,vasilev1989}
\begin{equation}
\frac{d}{\xi_N} \frac{\rho_S \xi_S}{\rho_N \xi_N } \lesssim 0.3, 
\label{JCres}
\end{equation}
where $\rho_S$  and $\xi_S$ are resistivity and coherence length of the S-film. Going beyond the one-dimensional approximation, shows \cite{bosboom2021} that as $d$ increases, the temperature  $T_{Th}$ actually decreases. This automatically leads to an increase in the effective size of the structure with respect to the $L_{ef}$ defined by expression (\ref{Leff}).
}

{2. The boundary condition (\ref{BCinf}) states that the current flowing through the N-layer of the SN composite electrode is negligible compared to the current that has entered the S-layer of the electrode, i.e.
\begin{equation}
\frac{d \rho_S}{d_S \rho_N} \ll (1+\gamma_{BM})^2, 
\label{cond2}
\end{equation}
where $d_S$ is a thickness of the S-film. A violation of this constraint should lead to the appearance of a dependence of $\varphi$ on the coordinate $x$ (see the discussion in point 4. below). 
}

{3. At any point of the SN boundary, the density of the supercurrent injected from normal metal through this interface into the S-layer should be less than the critical current density provided by the finite transparency of this boundary, i.e.
\begin{equation}
1+ \gamma_{BM}^{-1}  > \frac{d}{\zeta} \max \left[\frac{\xi_N}{L},1\right]
\end{equation}
}

{4. The last restriction comes from the suggestion that phase difference $\varphi$ in (\ref{EqInN}), (\ref{ksieff}) is independent on coordinate $x$ value. This condition holds well for $x\gg L/2$. Away from the N-NS boundary, the superconducting current is uniformly distributed across the thickness of the NS electrode. The current flows along the $0x$ direction, so the current lines do not cross the NS boundary. As a result, at $x \gg L/2$ the phase of the order parameter in each cross sections of the NS electrode turns out to be equal to the phase of the anomalous Green's functions in both N and S-layers, and the $\varphi$ turns out to be a value independent of $x$.
}

\section{Conclusion}

Thus, through analytical evaluations and numerical calculations, we have shown that the Thouless energy in Josephson SN-N-NS bridges strongly depends on the SN interface resistance and decreases by several times with its growth compared to the SNS sandwich-type contact. 
We also formulate the range of validity of our analytical results, thus giving the possible reasons for the deviation of an experimentally obtained $L_{ef}$ from our analytical result (\ref{Leff}). 

{The Thouless energy cannot be measured directly. 
Instead, the Thouless temperature can be determined by analyzing the shape characteristics of experimentally obtained $I_c(T)$ dependencies. The Thouless temperature is a point on the temperature axis of an $I_c(T)$ plot where 
the critical current reaches saturation, so that the first derivative at $T \lesssim T_{Th}$ is $dI_c/dT$ closed to zero.  The use of this method for the experimental determination of $T_{Th}$ requires measurements at rather low temperatures, which causes additional difficulties in finding $T_{Th}$. We have shown that the transition from the study of the dependence $I_c(T)$ to the determination of special points on the curves $dIc(T)/dT$ located at $T \approx E_{T}/2$ and $T \approx E_{T}$ allows us to estimate the value of $E_{T}$ in the range of temperatures more convenient for measurements.} 

{Our numerical calculations carried out in the framework of the one-dimensional SN-N-NS model of the Josephson SN-N-NS structure have shown that there is a reasonable correspondence between the value of the Thouless energy $E_T= 2 \pi T_{Th}$ determined in this way and the effective geometrical size of the structure $L_{ef}$ (see Eq.(\ref{Leff})) included in formula (\ref{ThGsm}). The estimate of $L_{ef}$ obtained in this way is important in determining the constraints to miniaturization of the size of superconductor devices for processing analog and digital signals. It also provides experimentalists with additional information on the relationship between such important technological parameters as the NS electrode spacing $L$, decay length $\xi_N$, and the parameter $\gamma_{BM}$, which characterizes the transparency of NS boundaries.}




\section{Acknowledgments}

We are grateful to A.I. Zaikin and A.A. Golubov for the discussion of the results obtained. A.N. thanks the support of the Foundation for the Development of Theoretical Physics and Mathematics "BASIS". The authors are grateful to MIPT Collective Use Center for providing facilities for samples fabrication.


\bibliography{Arxiv.bib}

\begin{thebibliography}{54}
\expandafter\ifx\csname natexlab\endcsname\relax\def\natexlab#1{#1}\fi
\expandafter\ifx\csname bibnamefont\endcsname\relax
  \def\bibnamefont#1{#1}\fi
\expandafter\ifx\csname bibfnamefont\endcsname\relax
  \def\bibfnamefont#1{#1}\fi
\expandafter\ifx\csname citenamefont\endcsname\relax
  \def\citenamefont#1{#1}\fi
\expandafter\ifx\csname url\endcsname\relax
  \def\url#1{\texttt{#1}}\fi
\expandafter\ifx\csname urlprefix\endcsname\relax\def\urlprefix{URL }\fi
\providecommand{\bibinfo}[2]{#2}
\providecommand{\eprint}[2][]{\url{#2}}

\bibitem[{\citenamefont{Kupriyanov et~al.}(1999)\citenamefont{Kupriyanov, Brinkman, Golubov, Siegel, and Rogalla}}]{brinkman1999}
\bibinfo{author}{\bibfnamefont{M.~Y.} \bibnamefont{Kupriyanov}}, \bibinfo{author}{\bibfnamefont{A.}~\bibnamefont{Brinkman}}, \bibinfo{author}{\bibfnamefont{A.~A.} \bibnamefont{Golubov}}, \bibinfo{author}{\bibfnamefont{M.}~\bibnamefont{Siegel}}, \bibnamefont{and} \bibinfo{author}{\bibfnamefont{H.}~\bibnamefont{Rogalla}}, \bibinfo{journal}{Physica C: Superconductivity and its Applications} \textbf{\bibinfo{volume}{326}}, \bibinfo{pages}{16} (\bibinfo{year}{1999}).

\bibitem[{\citenamefont{Holmes et~al.}(2013)\citenamefont{Holmes, Ripple, and Manheimer}}]{Holmes}
\bibinfo{author}{\bibfnamefont{D.~S.} \bibnamefont{Holmes}}, \bibinfo{author}{\bibfnamefont{A.~L.} \bibnamefont{Ripple}}, \bibnamefont{and} \bibinfo{author}{\bibfnamefont{M.~A.} \bibnamefont{Manheimer}}, \bibinfo{journal}{IEEE Trans. Appl. Supercond.} \textbf{\bibinfo{volume}{23}}, \bibinfo{pages}{1701610} (\bibinfo{year}{2013}).

\bibitem[{\citenamefont{Shelly et~al.}(2017)\citenamefont{Shelly, See, Ireland, Romans, and Williams}}]{shelly2017weak}
\bibinfo{author}{\bibfnamefont{C.~D.} \bibnamefont{Shelly}}, \bibinfo{author}{\bibfnamefont{P.}~\bibnamefont{See}}, \bibinfo{author}{\bibfnamefont{J.}~\bibnamefont{Ireland}}, \bibinfo{author}{\bibfnamefont{E.~J.} \bibnamefont{Romans}}, \bibnamefont{and} \bibinfo{author}{\bibfnamefont{J.~M.} \bibnamefont{Williams}}, \bibinfo{journal}{Superconductor Science and Technology} \textbf{\bibinfo{volume}{30}}, \bibinfo{pages}{095013} (\bibinfo{year}{2017}).

\bibitem[{\citenamefont{Tolpygo}(2016)}]{Tolp}
\bibinfo{author}{\bibfnamefont{S.~K.} \bibnamefont{Tolpygo}}, \bibinfo{journal}{Low Temp. Phys.} \textbf{\bibinfo{volume}{42}}, \bibinfo{pages}{361} (\bibinfo{year}{2016}).

\bibitem[{\citenamefont{Semenov et~al.}(2019)\citenamefont{Semenov, Polyakov, and Tolpygo}}]{semenov2019very}
\bibinfo{author}{\bibfnamefont{V.~K.} \bibnamefont{Semenov}}, \bibinfo{author}{\bibfnamefont{Y.~A.} \bibnamefont{Polyakov}}, \bibnamefont{and} \bibinfo{author}{\bibfnamefont{S.~K.} \bibnamefont{Tolpygo}}, \bibinfo{journal}{IEEE Transactions on Applied Superconductivity} \textbf{\bibinfo{volume}{29}}, \bibinfo{pages}{1} (\bibinfo{year}{2019}).

\bibitem[{\citenamefont{Collins et~al.}(2022)\citenamefont{Collins, Rose, and Casaburi}}]{collins2022superconducting}
\bibinfo{author}{\bibfnamefont{J.~A.} \bibnamefont{Collins}}, \bibinfo{author}{\bibfnamefont{C.~S.} \bibnamefont{Rose}}, \bibnamefont{and} \bibinfo{author}{\bibfnamefont{A.}~\bibnamefont{Casaburi}}, \bibinfo{journal}{IEEE Transactions on Applied Superconductivity} \textbf{\bibinfo{volume}{33}}, \bibinfo{pages}{1} (\bibinfo{year}{2022}).

\bibitem[{\citenamefont{Thompson et~al.}(2022)\citenamefont{Thompson, Castellanos-Beltran, Hopkins, Dresselhaus, and Benz}}]{thompson2022effects}
\bibinfo{author}{\bibfnamefont{M.~L.} \bibnamefont{Thompson}}, \bibinfo{author}{\bibfnamefont{M.}~\bibnamefont{Castellanos-Beltran}}, \bibinfo{author}{\bibfnamefont{P.~F.} \bibnamefont{Hopkins}}, \bibinfo{author}{\bibfnamefont{P.~D.} \bibnamefont{Dresselhaus}}, \bibnamefont{and} \bibinfo{author}{\bibfnamefont{S.~P.} \bibnamefont{Benz}}, \bibinfo{journal}{IEEE Transactions on Applied Superconductivity} \textbf{\bibinfo{volume}{33}}, \bibinfo{pages}{1} (\bibinfo{year}{2022}).

\bibitem[{\citenamefont{Tolpygo et~al.}(2023)\citenamefont{Tolpygo, Rastogi, Weir, Golden, and Bolkhovsky}}]{tolpygo2023development}
\bibinfo{author}{\bibfnamefont{S.~K.} \bibnamefont{Tolpygo}}, \bibinfo{author}{\bibfnamefont{R.}~\bibnamefont{Rastogi}}, \bibinfo{author}{\bibfnamefont{T.}~\bibnamefont{Weir}}, \bibinfo{author}{\bibfnamefont{E.~B.} \bibnamefont{Golden}}, \bibnamefont{and} \bibinfo{author}{\bibfnamefont{V.}~\bibnamefont{Bolkhovsky}}, \bibinfo{journal}{arXiv preprint arXiv:2312.13475}  (\bibinfo{year}{2023}).

\bibitem[{\citenamefont{Tolpygo et~al.}(2024)\citenamefont{Tolpygo, Rastogi, Weir, Golden, and Bolkhovsky}}]{10431604}
\bibinfo{author}{\bibfnamefont{S.~K.} \bibnamefont{Tolpygo}}, \bibinfo{author}{\bibfnamefont{R.}~\bibnamefont{Rastogi}}, \bibinfo{author}{\bibfnamefont{T.}~\bibnamefont{Weir}}, \bibinfo{author}{\bibfnamefont{E.~B.} \bibnamefont{Golden}}, \bibnamefont{and} \bibinfo{author}{\bibfnamefont{V.}~\bibnamefont{Bolkhovsky}}, \bibinfo{journal}{IEEE Transactions on Applied Superconductivity} \textbf{\bibinfo{volume}{34}}, \bibinfo{pages}{1} (\bibinfo{year}{2024}).

\bibitem[{\citenamefont{Soloviev et~al.}(2021)\citenamefont{Soloviev, Bakurskiy, Ruzhickiy, Klenov, Kupriyanov, Golubov, Skryabina, and Stolyarov}}]{soloviev2021miniaturization}
\bibinfo{author}{\bibfnamefont{I.}~\bibnamefont{Soloviev}}, \bibinfo{author}{\bibfnamefont{S.}~\bibnamefont{Bakurskiy}}, \bibinfo{author}{\bibfnamefont{V.}~\bibnamefont{Ruzhickiy}}, \bibinfo{author}{\bibfnamefont{N.}~\bibnamefont{Klenov}}, \bibinfo{author}{\bibfnamefont{M.~Y.} \bibnamefont{Kupriyanov}}, \bibinfo{author}{\bibfnamefont{A.}~\bibnamefont{Golubov}}, \bibinfo{author}{\bibfnamefont{O.}~\bibnamefont{Skryabina}}, \bibnamefont{and} \bibinfo{author}{\bibfnamefont{V.}~\bibnamefont{Stolyarov}}, \bibinfo{journal}{Physical review applied} \textbf{\bibinfo{volume}{16}}, \bibinfo{pages}{044060} (\bibinfo{year}{2021}).

\bibitem[{\citenamefont{Ruzhickiy et~al.}(2023)\citenamefont{Ruzhickiy, Bakurskiy, Kupriyanov, Klenov, Soloviev, Stolyarov, and Golubov}}]{nano13121873}
\bibinfo{author}{\bibfnamefont{V.}~\bibnamefont{Ruzhickiy}}, \bibinfo{author}{\bibfnamefont{S.}~\bibnamefont{Bakurskiy}}, \bibinfo{author}{\bibfnamefont{M.}~\bibnamefont{Kupriyanov}}, \bibinfo{author}{\bibfnamefont{N.}~\bibnamefont{Klenov}}, \bibinfo{author}{\bibfnamefont{I.}~\bibnamefont{Soloviev}}, \bibinfo{author}{\bibfnamefont{V.}~\bibnamefont{Stolyarov}}, \bibnamefont{and} \bibinfo{author}{\bibfnamefont{A.}~\bibnamefont{Golubov}}, \bibinfo{journal}{Nanomaterials} \textbf{\bibinfo{volume}{13}} (\bibinfo{year}{2023}).

\bibitem[{\citenamefont{Likharev}(1979)}]{likharev1979superconducting}
\bibinfo{author}{\bibfnamefont{K.}~\bibnamefont{Likharev}}, \bibinfo{journal}{Reviews of Modern Physics} \textbf{\bibinfo{volume}{51}}, \bibinfo{pages}{101} (\bibinfo{year}{1979}).

\bibitem[{\citenamefont{Golubov et~al.}(2004)\citenamefont{Golubov, Kupriyanov, and Il’Ichev}}]{golubov2004current}
\bibinfo{author}{\bibfnamefont{A.~A.} \bibnamefont{Golubov}}, \bibinfo{author}{\bibfnamefont{M.~Y.} \bibnamefont{Kupriyanov}}, \bibnamefont{and} \bibinfo{author}{\bibfnamefont{E.}~\bibnamefont{Il’Ichev}}, \bibinfo{journal}{Reviews of modern physics} \textbf{\bibinfo{volume}{76}}, \bibinfo{pages}{411} (\bibinfo{year}{2004}).

\bibitem[{\citenamefont{Likharev}(1976)}]{likharev1976sov}
\bibinfo{author}{\bibfnamefont{K.}~\bibnamefont{Likharev}}, \bibinfo{journal}{Sov. Tech. Phys. Lett} \textbf{\bibinfo{volume}{2}}, \bibinfo{pages}{12} (\bibinfo{year}{1976}).

\bibitem[{\citenamefont{Ivanov et~al.}(1981)\citenamefont{Ivanov, Kupriyanov, Likharev, Meriakri, and Snigirev}}]{likharev1981boundary}
\bibinfo{author}{\bibfnamefont{Z.~G.} \bibnamefont{Ivanov}}, \bibinfo{author}{\bibfnamefont{M.~Y.} \bibnamefont{Kupriyanov}}, \bibinfo{author}{\bibfnamefont{K.~K.} \bibnamefont{Likharev}}, \bibinfo{author}{\bibfnamefont{S.~V.} \bibnamefont{Meriakri}}, \bibnamefont{and} \bibinfo{author}{\bibfnamefont{O.~V.} \bibnamefont{Snigirev}}, \bibinfo{journal}{Fizika Nizkikh Temperatur} \textbf{\bibinfo{volume}{7}}, \bibinfo{pages}{560} (\bibinfo{year}{1981}), \bibinfo{note}{[Sov. J. Low. Temp. Phys. 7, 274-281 (1981)]}.

\bibitem[{\citenamefont{Zaikin and Zharkov}(1981)}]{zaikin1981theory}
\bibinfo{author}{\bibfnamefont{A.}~\bibnamefont{Zaikin}} \bibnamefont{and} \bibinfo{author}{\bibfnamefont{G.}~\bibnamefont{Zharkov}}, \bibinfo{journal}{Soviet Journal of Low Temperature Physics} \textbf{\bibinfo{volume}{7}}, \bibinfo{pages}{184} (\bibinfo{year}{1981}).

\bibitem[{\citenamefont{Dubos et~al.}(2001)\citenamefont{Dubos, Courtois, Pannetier, Wilhelm, Zaikin, and Sch{\"o}n}}]{dubos2001josephson}
\bibinfo{author}{\bibfnamefont{P.}~\bibnamefont{Dubos}}, \bibinfo{author}{\bibfnamefont{H.}~\bibnamefont{Courtois}}, \bibinfo{author}{\bibfnamefont{B.}~\bibnamefont{Pannetier}}, \bibinfo{author}{\bibfnamefont{F.}~\bibnamefont{Wilhelm}}, \bibinfo{author}{\bibfnamefont{A.}~\bibnamefont{Zaikin}}, \bibnamefont{and} \bibinfo{author}{\bibfnamefont{G.}~\bibnamefont{Sch{\"o}n}}, \bibinfo{journal}{Physical Review B} \textbf{\bibinfo{volume}{63}}, \bibinfo{pages}{064502} (\bibinfo{year}{2001}).

\bibitem[{\citenamefont{Hammer et~al.}(2007)\citenamefont{Hammer, Cuevas, Bergeret, and Belzig}}]{hammer2007density}
\bibinfo{author}{\bibfnamefont{J.}~\bibnamefont{Hammer}}, \bibinfo{author}{\bibfnamefont{J.~C.} \bibnamefont{Cuevas}}, \bibinfo{author}{\bibfnamefont{F.}~\bibnamefont{Bergeret}}, \bibnamefont{and} \bibinfo{author}{\bibfnamefont{W.}~\bibnamefont{Belzig}}, \bibinfo{journal}{Physical Review B} \textbf{\bibinfo{volume}{76}}, \bibinfo{pages}{064514} (\bibinfo{year}{2007}).

\bibitem[{\citenamefont{Marychev and Vodolazov}(2020)}]{marychev2020josephson}
\bibinfo{author}{\bibfnamefont{P.~M.} \bibnamefont{Marychev}} \bibnamefont{and} \bibinfo{author}{\bibfnamefont{D.~Y.} \bibnamefont{Vodolazov}}, \bibinfo{journal}{Beilstein Journal of Nanotechnology} \textbf{\bibinfo{volume}{11}}, \bibinfo{pages}{858} (\bibinfo{year}{2020}).

\bibitem[{\citenamefont{Giazotto et~al.}(2006)\citenamefont{Giazotto, Heikkil{\"a}, Luukanen, Savin, and Pekola}}]{giazotto2006opportunities}
\bibinfo{author}{\bibfnamefont{F.}~\bibnamefont{Giazotto}}, \bibinfo{author}{\bibfnamefont{T.~T.} \bibnamefont{Heikkil{\"a}}}, \bibinfo{author}{\bibfnamefont{A.}~\bibnamefont{Luukanen}}, \bibinfo{author}{\bibfnamefont{A.~M.} \bibnamefont{Savin}}, \bibnamefont{and} \bibinfo{author}{\bibfnamefont{J.~P.} \bibnamefont{Pekola}}, \bibinfo{journal}{Reviews of Modern Physics} \textbf{\bibinfo{volume}{78}}, \bibinfo{pages}{217} (\bibinfo{year}{2006}).

\bibitem[{\citenamefont{Angers et~al.}(2008)\citenamefont{Angers, Chiodi, Montambaux, Ferrier, Gu{\'e}ron, Bouchiat, and Cuevas}}]{Angers_2008}
\bibinfo{author}{\bibfnamefont{L.}~\bibnamefont{Angers}}, \bibinfo{author}{\bibfnamefont{F.}~\bibnamefont{Chiodi}}, \bibinfo{author}{\bibfnamefont{G.}~\bibnamefont{Montambaux}}, \bibinfo{author}{\bibfnamefont{M.}~\bibnamefont{Ferrier}}, \bibinfo{author}{\bibfnamefont{S.}~\bibnamefont{Gu{\'e}ron}}, \bibinfo{author}{\bibfnamefont{H.}~\bibnamefont{Bouchiat}}, \bibnamefont{and} \bibinfo{author}{\bibfnamefont{J.~C.} \bibnamefont{Cuevas}}, \bibinfo{journal}{Physical Review B} \textbf{\bibinfo{volume}{77}}, \bibinfo{pages}{165408} (\bibinfo{year}{2008}).

\bibitem[{\citenamefont{Courtois et~al.}(2008)\citenamefont{Courtois, Meschke, Peltonen, and Pekola}}]{courtois2008origin}
\bibinfo{author}{\bibfnamefont{H.}~\bibnamefont{Courtois}}, \bibinfo{author}{\bibfnamefont{M.}~\bibnamefont{Meschke}}, \bibinfo{author}{\bibfnamefont{J.}~\bibnamefont{Peltonen}}, \bibnamefont{and} \bibinfo{author}{\bibfnamefont{J.~P.} \bibnamefont{Pekola}}, \bibinfo{journal}{Physical review letters} \textbf{\bibinfo{volume}{101}}, \bibinfo{pages}{067002} (\bibinfo{year}{2008}).

\bibitem[{\citenamefont{Carillo et~al.}(2008)\citenamefont{Carillo, Born, Pellegrini, Tafuri, Biasiol, Sorba, and Beltram}}]{PhysRevB.78.052506}
\bibinfo{author}{\bibfnamefont{F.}~\bibnamefont{Carillo}}, \bibinfo{author}{\bibfnamefont{D.}~\bibnamefont{Born}}, \bibinfo{author}{\bibfnamefont{V.}~\bibnamefont{Pellegrini}}, \bibinfo{author}{\bibfnamefont{F.}~\bibnamefont{Tafuri}}, \bibinfo{author}{\bibfnamefont{G.}~\bibnamefont{Biasiol}}, \bibinfo{author}{\bibfnamefont{L.}~\bibnamefont{Sorba}}, \bibnamefont{and} \bibinfo{author}{\bibfnamefont{F.}~\bibnamefont{Beltram}}, \bibinfo{journal}{Phys. Rev. B} \textbf{\bibinfo{volume}{78}}, \bibinfo{pages}{052506} (\bibinfo{year}{2008}).

\bibitem[{\citenamefont{García and Giazotto}(2009)}]{García_2009}
\bibinfo{author}{\bibfnamefont{C.~P.} \bibnamefont{García}} \bibnamefont{and} \bibinfo{author}{\bibfnamefont{F.}~\bibnamefont{Giazotto}}, \bibinfo{journal}{Appl. Phys. Lett.} \textbf{\bibinfo{volume}{94}}, \bibinfo{pages}{132508} (\bibinfo{year}{2009}).

\bibitem[{\citenamefont{Giazotto et~al.}(2010)\citenamefont{Giazotto, Peltonen, Meschke, and Pekola}}]{giazotto2010superconducting}
\bibinfo{author}{\bibfnamefont{F.}~\bibnamefont{Giazotto}}, \bibinfo{author}{\bibfnamefont{J.~T.} \bibnamefont{Peltonen}}, \bibinfo{author}{\bibfnamefont{M.}~\bibnamefont{Meschke}}, \bibnamefont{and} \bibinfo{author}{\bibfnamefont{J.~P.} \bibnamefont{Pekola}}, \bibinfo{journal}{Nature Physics} \textbf{\bibinfo{volume}{6}}, \bibinfo{pages}{254} (\bibinfo{year}{2010}).

\bibitem[{\citenamefont{Frielinghaus et~al.}(2010)\citenamefont{Frielinghaus, Batov, Weides, Kohlstedt, Calarco, and Sch{\"a}pers}}]{frielinghaus2010josephson}
\bibinfo{author}{\bibfnamefont{R.}~\bibnamefont{Frielinghaus}}, \bibinfo{author}{\bibfnamefont{I.}~\bibnamefont{Batov}}, \bibinfo{author}{\bibfnamefont{M.}~\bibnamefont{Weides}}, \bibinfo{author}{\bibfnamefont{H.}~\bibnamefont{Kohlstedt}}, \bibinfo{author}{\bibfnamefont{R.}~\bibnamefont{Calarco}}, \bibnamefont{and} \bibinfo{author}{\bibfnamefont{T.}~\bibnamefont{Sch{\"a}pers}}, \bibinfo{journal}{Applied physics letters} \textbf{\bibinfo{volume}{96}} (\bibinfo{year}{2010}).

\bibitem[{\citenamefont{Jung et~al.}(2011)\citenamefont{Jung, Noh, Doh, Song, Chong, Choi, Yoo, Seo, Kim, Woo et~al.}}]{jung2011superconducting}
\bibinfo{author}{\bibfnamefont{M.}~\bibnamefont{Jung}}, \bibinfo{author}{\bibfnamefont{H.}~\bibnamefont{Noh}}, \bibinfo{author}{\bibfnamefont{Y.-J.} \bibnamefont{Doh}}, \bibinfo{author}{\bibfnamefont{W.}~\bibnamefont{Song}}, \bibinfo{author}{\bibfnamefont{Y.}~\bibnamefont{Chong}}, \bibinfo{author}{\bibfnamefont{M.-S.} \bibnamefont{Choi}}, \bibinfo{author}{\bibfnamefont{Y.}~\bibnamefont{Yoo}}, \bibinfo{author}{\bibfnamefont{K.}~\bibnamefont{Seo}}, \bibinfo{author}{\bibfnamefont{N.}~\bibnamefont{Kim}}, \bibinfo{author}{\bibfnamefont{B.-C.} \bibnamefont{Woo}}, \bibnamefont{et~al.}, \bibinfo{journal}{Acs Nano} \textbf{\bibinfo{volume}{5}}, \bibinfo{pages}{2271} (\bibinfo{year}{2011}).

\bibitem[{\citenamefont{Golikova et~al.}(2012)\citenamefont{Golikova, Hubler, Beckmann, Klenov, Bakurskiy, Kupriyanov, Batov, and Ryazanov}}]{batov2012critical3362877}
\bibinfo{author}{\bibfnamefont{T.~E.} \bibnamefont{Golikova}}, \bibinfo{author}{\bibfnamefont{F.}~\bibnamefont{Hubler}}, \bibinfo{author}{\bibfnamefont{D.}~\bibnamefont{Beckmann}}, \bibinfo{author}{\bibfnamefont{N.~V.} \bibnamefont{Klenov}}, \bibinfo{author}{\bibfnamefont{S.~V.} \bibnamefont{Bakurskiy}}, \bibinfo{author}{\bibfnamefont{M.~Y.} \bibnamefont{Kupriyanov}}, \bibinfo{author}{\bibfnamefont{I.~E.} \bibnamefont{Batov}}, \bibnamefont{and} \bibinfo{author}{\bibfnamefont{V.~V.} \bibnamefont{Ryazanov}}, \bibinfo{journal}{JETP Letters} \textbf{\bibinfo{volume}{96}}, \bibinfo{pages}{668} (\bibinfo{year}{2012}).

\bibitem[{\citenamefont{Golikova et~al.}(2014)\citenamefont{Golikova, Wolf, Beckmann, Batov, Bobkova, Bobkov, and Ryazanov}}]{PhysRevB.89.104507}
\bibinfo{author}{\bibfnamefont{T.~E.} \bibnamefont{Golikova}}, \bibinfo{author}{\bibfnamefont{M.~J.} \bibnamefont{Wolf}}, \bibinfo{author}{\bibfnamefont{D.}~\bibnamefont{Beckmann}}, \bibinfo{author}{\bibfnamefont{I.~E.} \bibnamefont{Batov}}, \bibinfo{author}{\bibfnamefont{I.~V.} \bibnamefont{Bobkova}}, \bibinfo{author}{\bibfnamefont{A.~M.} \bibnamefont{Bobkov}}, \bibnamefont{and} \bibinfo{author}{\bibfnamefont{V.~V.} \bibnamefont{Ryazanov}}, \bibinfo{journal}{Phys. Rev. B} \textbf{\bibinfo{volume}{89}}, \bibinfo{pages}{104507} (\bibinfo{year}{2014}).

\bibitem[{\citenamefont{Pekola}(2015)}]{pekola2015towards}
\bibinfo{author}{\bibfnamefont{J.~P.} \bibnamefont{Pekola}}, \bibinfo{journal}{Nature physics} \textbf{\bibinfo{volume}{11}}, \bibinfo{pages}{118} (\bibinfo{year}{2015}).

\bibitem[{\citenamefont{Paajaste et~al.}(2015)\citenamefont{Paajaste, Amado, Roddaro, Bergeret, Ercolani, Sorba, and Giazotto}}]{paajaste2015pb}
\bibinfo{author}{\bibfnamefont{J.}~\bibnamefont{Paajaste}}, \bibinfo{author}{\bibfnamefont{M.}~\bibnamefont{Amado}}, \bibinfo{author}{\bibfnamefont{S.}~\bibnamefont{Roddaro}}, \bibinfo{author}{\bibfnamefont{F.}~\bibnamefont{Bergeret}}, \bibinfo{author}{\bibfnamefont{D.}~\bibnamefont{Ercolani}}, \bibinfo{author}{\bibfnamefont{L.}~\bibnamefont{Sorba}}, \bibnamefont{and} \bibinfo{author}{\bibfnamefont{F.}~\bibnamefont{Giazotto}}, \bibinfo{journal}{Nano letters} \textbf{\bibinfo{volume}{15}}, \bibinfo{pages}{1803} (\bibinfo{year}{2015}).

\bibitem[{\citenamefont{De~Cecco et~al.}(2016)\citenamefont{De~Cecco, Le~Calvez, Sac{\'e}p{\'e}, Winkelmann, and Courtois}}]{de2016interplay}
\bibinfo{author}{\bibfnamefont{A.}~\bibnamefont{De~Cecco}}, \bibinfo{author}{\bibfnamefont{K.}~\bibnamefont{Le~Calvez}}, \bibinfo{author}{\bibfnamefont{B.}~\bibnamefont{Sac{\'e}p{\'e}}}, \bibinfo{author}{\bibfnamefont{C.}~\bibnamefont{Winkelmann}}, \bibnamefont{and} \bibinfo{author}{\bibfnamefont{H.}~\bibnamefont{Courtois}}, \bibinfo{journal}{Physical Review B} \textbf{\bibinfo{volume}{93}}, \bibinfo{pages}{180505} (\bibinfo{year}{2016}).

\bibitem[{\citenamefont{Jabdaraghi et~al.}(2016)\citenamefont{Jabdaraghi, Peltonen, Saira, and Pekola}}]{jabdaraghi2016low}
\bibinfo{author}{\bibfnamefont{R.~N.} \bibnamefont{Jabdaraghi}}, \bibinfo{author}{\bibfnamefont{J.}~\bibnamefont{Peltonen}}, \bibinfo{author}{\bibfnamefont{O.-P.} \bibnamefont{Saira}}, \bibnamefont{and} \bibinfo{author}{\bibfnamefont{J.}~\bibnamefont{Pekola}}, \bibinfo{journal}{Applied Physics Letters} \textbf{\bibinfo{volume}{108}} (\bibinfo{year}{2016}).

\bibitem[{\citenamefont{Skryabina et~al.}(2017)\citenamefont{Skryabina, Egorov, Goncharova, Klimenko, Kozlov, Ryazanov, Bakurskiy, Kupriyanov, Golubov, Napolskii et~al.}}]{skryabina2017josephson}
\bibinfo{author}{\bibfnamefont{O.}~\bibnamefont{Skryabina}}, \bibinfo{author}{\bibfnamefont{S.}~\bibnamefont{Egorov}}, \bibinfo{author}{\bibfnamefont{A.}~\bibnamefont{Goncharova}}, \bibinfo{author}{\bibfnamefont{A.}~\bibnamefont{Klimenko}}, \bibinfo{author}{\bibfnamefont{S.}~\bibnamefont{Kozlov}}, \bibinfo{author}{\bibfnamefont{V.}~\bibnamefont{Ryazanov}}, \bibinfo{author}{\bibfnamefont{S.}~\bibnamefont{Bakurskiy}}, \bibinfo{author}{\bibfnamefont{M.~Y.} \bibnamefont{Kupriyanov}}, \bibinfo{author}{\bibfnamefont{A.}~\bibnamefont{Golubov}}, \bibinfo{author}{\bibfnamefont{K.}~\bibnamefont{Napolskii}}, \bibnamefont{et~al.}, \bibinfo{journal}{Applied physics letters} \textbf{\bibinfo{volume}{110}} (\bibinfo{year}{2017}).

\bibitem[{\citenamefont{Kim et~al.}(2017)\citenamefont{Kim, Kim, Yang, Peng, Yu, and Doh}}]{kim2017strong}
\bibinfo{author}{\bibfnamefont{B.-K.} \bibnamefont{Kim}}, \bibinfo{author}{\bibfnamefont{H.-S.} \bibnamefont{Kim}}, \bibinfo{author}{\bibfnamefont{Y.}~\bibnamefont{Yang}}, \bibinfo{author}{\bibfnamefont{X.}~\bibnamefont{Peng}}, \bibinfo{author}{\bibfnamefont{D.}~\bibnamefont{Yu}}, \bibnamefont{and} \bibinfo{author}{\bibfnamefont{Y.-J.} \bibnamefont{Doh}}, \bibinfo{journal}{ACS nano} \textbf{\bibinfo{volume}{11}}, \bibinfo{pages}{221} (\bibinfo{year}{2017}).

\bibitem[{\citenamefont{Shishkin et~al.}(2020)\citenamefont{Shishkin, Skryabina, Gurtovoi, Dizhur, Faley, Golubov, and Stolyarov}}]{shishkin2020planar}
\bibinfo{author}{\bibfnamefont{A.}~\bibnamefont{Shishkin}}, \bibinfo{author}{\bibfnamefont{O.}~\bibnamefont{Skryabina}}, \bibinfo{author}{\bibfnamefont{V.}~\bibnamefont{Gurtovoi}}, \bibinfo{author}{\bibfnamefont{S.}~\bibnamefont{Dizhur}}, \bibinfo{author}{\bibfnamefont{M.}~\bibnamefont{Faley}}, \bibinfo{author}{\bibfnamefont{A.}~\bibnamefont{Golubov}}, \bibnamefont{and} \bibinfo{author}{\bibfnamefont{V.}~\bibnamefont{Stolyarov}}, \bibinfo{journal}{Superconductor science and technology} \textbf{\bibinfo{volume}{33}}, \bibinfo{pages}{065005} (\bibinfo{year}{2020}).

\bibitem[{\citenamefont{Zhang et~al.}(2020)\citenamefont{Zhang, Jalil, Tse, Kölzer, Rosenbach, Valencia, Luysberg, Mikulics, Panaitov, Grützmacher et~al.}}]{Zhang2020}
\bibinfo{author}{\bibfnamefont{J.}~\bibnamefont{Zhang}}, \bibinfo{author}{\bibfnamefont{A.~R.} \bibnamefont{Jalil}}, \bibinfo{author}{\bibfnamefont{P.-L.} \bibnamefont{Tse}}, \bibinfo{author}{\bibfnamefont{J.}~\bibnamefont{Kölzer}}, \bibinfo{author}{\bibfnamefont{D.}~\bibnamefont{Rosenbach}}, \bibinfo{author}{\bibfnamefont{H.}~\bibnamefont{Valencia}}, \bibinfo{author}{\bibfnamefont{M.}~\bibnamefont{Luysberg}}, \bibinfo{author}{\bibfnamefont{M.}~\bibnamefont{Mikulics}}, \bibinfo{author}{\bibfnamefont{G.}~\bibnamefont{Panaitov}}, \bibinfo{author}{\bibfnamefont{D.}~\bibnamefont{Grützmacher}}, \bibnamefont{et~al.}, \bibinfo{journal}{Annalen der Physik} \textbf{\bibinfo{volume}{532}}, \bibinfo{pages}{2000273} (\bibinfo{year}{2020}).

\bibitem[{\citenamefont{Murani et~al.}(2020)\citenamefont{Murani, Sengupta, Kasumov, Deblock, Celle, Simonato, Bouchiat, and Gu{\'e}ron}}]{murani2020long}
\bibinfo{author}{\bibfnamefont{A.}~\bibnamefont{Murani}}, \bibinfo{author}{\bibfnamefont{S.}~\bibnamefont{Sengupta}}, \bibinfo{author}{\bibfnamefont{A.}~\bibnamefont{Kasumov}}, \bibinfo{author}{\bibfnamefont{R.}~\bibnamefont{Deblock}}, \bibinfo{author}{\bibfnamefont{C.}~\bibnamefont{Celle}}, \bibinfo{author}{\bibfnamefont{J.}~\bibnamefont{Simonato}}, \bibinfo{author}{\bibfnamefont{H.}~\bibnamefont{Bouchiat}}, \bibnamefont{and} \bibinfo{author}{\bibfnamefont{S.}~\bibnamefont{Gu{\'e}ron}}, \bibinfo{journal}{Physical Review B} \textbf{\bibinfo{volume}{102}}, \bibinfo{pages}{214506} (\bibinfo{year}{2020}).

\bibitem[{\citenamefont{Skryabina et~al.}(2021)\citenamefont{Skryabina, Bakurskiy, Shishkin, Klimenko, Napolskii, Klenov, Soloviev, Ryazanov, Golubov, Roditchev et~al.}}]{skryabina2021environment}
\bibinfo{author}{\bibfnamefont{O.}~\bibnamefont{Skryabina}}, \bibinfo{author}{\bibfnamefont{S.}~\bibnamefont{Bakurskiy}}, \bibinfo{author}{\bibfnamefont{A.}~\bibnamefont{Shishkin}}, \bibinfo{author}{\bibfnamefont{A.}~\bibnamefont{Klimenko}}, \bibinfo{author}{\bibfnamefont{K.}~\bibnamefont{Napolskii}}, \bibinfo{author}{\bibfnamefont{N.}~\bibnamefont{Klenov}}, \bibinfo{author}{\bibfnamefont{I.}~\bibnamefont{Soloviev}}, \bibinfo{author}{\bibfnamefont{V.}~\bibnamefont{Ryazanov}}, \bibinfo{author}{\bibfnamefont{A.}~\bibnamefont{Golubov}}, \bibinfo{author}{\bibfnamefont{D.}~\bibnamefont{Roditchev}}, \bibnamefont{et~al.}, \bibinfo{journal}{Scientific reports} \textbf{\bibinfo{volume}{11}}, \bibinfo{pages}{15274} (\bibinfo{year}{2021}).

\bibitem[{\citenamefont{Golod et~al.}(2021)\citenamefont{Golod, Hovhannisyan, Kapran, Dremov, Stolyarov, and Krasnov}}]{golod2021reconfigurable}
\bibinfo{author}{\bibfnamefont{T.}~\bibnamefont{Golod}}, \bibinfo{author}{\bibfnamefont{R.~A.} \bibnamefont{Hovhannisyan}}, \bibinfo{author}{\bibfnamefont{O.~M.} \bibnamefont{Kapran}}, \bibinfo{author}{\bibfnamefont{V.~V.} \bibnamefont{Dremov}}, \bibinfo{author}{\bibfnamefont{V.~S.} \bibnamefont{Stolyarov}}, \bibnamefont{and} \bibinfo{author}{\bibfnamefont{V.~M.} \bibnamefont{Krasnov}}, \bibinfo{journal}{Nano Letters} \textbf{\bibinfo{volume}{21}}, \bibinfo{pages}{5240} (\bibinfo{year}{2021}).

\bibitem[{\citenamefont{Sotnichuk et~al.}(2022)\citenamefont{Sotnichuk, Skryabina, Shishkin, Bakurskiy, Kupriyanov, Stolyarov, and Napolskii}}]{sotnichuk2022long}
\bibinfo{author}{\bibfnamefont{S.~V.} \bibnamefont{Sotnichuk}}, \bibinfo{author}{\bibfnamefont{O.~V.} \bibnamefont{Skryabina}}, \bibinfo{author}{\bibfnamefont{A.~G.} \bibnamefont{Shishkin}}, \bibinfo{author}{\bibfnamefont{S.~V.} \bibnamefont{Bakurskiy}}, \bibinfo{author}{\bibfnamefont{M.~Y.} \bibnamefont{Kupriyanov}}, \bibinfo{author}{\bibfnamefont{V.~S.} \bibnamefont{Stolyarov}}, \bibnamefont{and} \bibinfo{author}{\bibfnamefont{K.~S.} \bibnamefont{Napolskii}}, \bibinfo{journal}{ACS Applied Nano Materials} \textbf{\bibinfo{volume}{5}}, \bibinfo{pages}{17059} (\bibinfo{year}{2022}).

\bibitem[{\citenamefont{Babich et~al.}(2023)\citenamefont{Babich, Kudriashov, Baranov, and Stolyarov}}]{babich2023limitations}
\bibinfo{author}{\bibfnamefont{I.}~\bibnamefont{Babich}}, \bibinfo{author}{\bibfnamefont{A.}~\bibnamefont{Kudriashov}}, \bibinfo{author}{\bibfnamefont{D.}~\bibnamefont{Baranov}}, \bibnamefont{and} \bibinfo{author}{\bibfnamefont{V.~S.} \bibnamefont{Stolyarov}}, \bibinfo{journal}{Nano Letters} \textbf{\bibinfo{volume}{23}}, \bibinfo{pages}{6713} (\bibinfo{year}{2023}).

\bibitem[{\citenamefont{Edwards and Thouless}(1972)}]{Edwards_1972}
\bibinfo{author}{\bibfnamefont{J.~T.} \bibnamefont{Edwards}} \bibnamefont{and} \bibinfo{author}{\bibfnamefont{D.~J.} \bibnamefont{Thouless}}, \bibinfo{journal}{Journal of Physics C: Solid State Physics} \textbf{\bibinfo{volume}{5}}, \bibinfo{pages}{807} (\bibinfo{year}{1972}).

\bibitem[{\citenamefont{Thouless}(1977)}]{PhysRevLett.39.1167}
\bibinfo{author}{\bibfnamefont{D.~J.} \bibnamefont{Thouless}}, \bibinfo{journal}{Phys. Rev. Lett.} \textbf{\bibinfo{volume}{39}}, \bibinfo{pages}{1167} (\bibinfo{year}{1977}).

\bibitem[{\citenamefont{Thouless}(1974)}]{THOULESS197493}
\bibinfo{author}{\bibfnamefont{D.}~\bibnamefont{Thouless}}, \bibinfo{journal}{Physics Reports} \textbf{\bibinfo{volume}{13}}, \bibinfo{pages}{93} (\bibinfo{year}{1974}), ISSN \bibinfo{issn}{0370-1573}.

\bibitem[{\citenamefont{Altland et~al.}(1996)\citenamefont{Altland, Gefen, and Montambaux}}]{PhysRevLett.76.1130}
\bibinfo{author}{\bibfnamefont{A.}~\bibnamefont{Altland}}, \bibinfo{author}{\bibfnamefont{Y.}~\bibnamefont{Gefen}}, \bibnamefont{and} \bibinfo{author}{\bibfnamefont{G.}~\bibnamefont{Montambaux}}, \bibinfo{journal}{Phys. Rev. Lett.} \textbf{\bibinfo{volume}{76}}, \bibinfo{pages}{1130} (\bibinfo{year}{1996}).

\bibitem[{\citenamefont{Usadel}(1970)}]{usadel1970}
\bibinfo{author}{\bibfnamefont{K.~D.} \bibnamefont{Usadel}}, \bibinfo{journal}{Physical Review Letters} \textbf{\bibinfo{volume}{25}}, \bibinfo{pages}{507} (\bibinfo{year}{1970}).

\bibitem[{\citenamefont{Golubov et~al.}(2005)\citenamefont{Golubov, Kupriyanov, and Siegel}}]{siegel2005density}
\bibinfo{author}{\bibfnamefont{A.~A.} \bibnamefont{Golubov}}, \bibinfo{author}{\bibfnamefont{M.~Y.} \bibnamefont{Kupriyanov}}, \bibnamefont{and} \bibinfo{author}{\bibfnamefont{M.}~\bibnamefont{Siegel}}, \bibinfo{journal}{Pis’ma v Zhurnal Éksperimental’noœ i Teoreticheskoœ Fiziki} \textbf{\bibinfo{volume}{81}}, \bibinfo{pages}{217} (\bibinfo{year}{2005}), \bibinfo{note}{[JETP Letters, 81 180-184 (2005)]}.

\bibitem[{\citenamefont{Likharev and Iakobson}(1975)}]{likharev1975steady}
\bibinfo{author}{\bibfnamefont{K.}~\bibnamefont{Likharev}} \bibnamefont{and} \bibinfo{author}{\bibfnamefont{L.}~\bibnamefont{Iakobson}}, \bibinfo{journal}{Zhurnal Tekhnicheskoi Fiziki} \textbf{\bibinfo{volume}{45}}, \bibinfo{pages}{1503} (\bibinfo{year}{1975}), \bibinfo{note}{[Sov. Phys. -Tech. Phys. 20, 950 (1975)]}.

\bibitem[{\citenamefont{Kupriyanov and Lukichev}(1982)}]{lukichev1982}
\bibinfo{author}{\bibfnamefont{M.~Y.} \bibnamefont{Kupriyanov}} \bibnamefont{and} \bibinfo{author}{\bibfnamefont{V.~F.} \bibnamefont{Lukichev}}, \bibinfo{journal}{Fizika Nizkikh Temperatur} \textbf{\bibinfo{volume}{8}}, \bibinfo{pages}{1045} (\bibinfo{year}{1982}), \bibinfo{note}{[Soviet Journal of Low Temperature Physics 8, 526-529 (1982)]}.

\bibitem[{Eli()}]{ElistratovaFuture}
\bibinfo{note}{Fabrication process details along with the full experimantal data for these series of the junctions will be publushed elsewhere soon}.

\bibitem[{\citenamefont{Kupriyanov et~al.}(1986)\citenamefont{Kupriyanov, Lukichev, and Orlikovskii}}]{orlikovskii1986}
\bibinfo{author}{\bibfnamefont{M.~Y.} \bibnamefont{Kupriyanov}}, \bibinfo{author}{\bibfnamefont{V.~F.} \bibnamefont{Lukichev}}, \bibnamefont{and} \bibinfo{author}{\bibfnamefont{A.~A.} \bibnamefont{Orlikovskii}}, \bibinfo{journal}{Microelectronics} \textbf{\bibinfo{volume}{15}}, \bibinfo{pages}{328} (\bibinfo{year}{1986}), \bibinfo{note}{[Soviet Microelectronics, 15, No4, 185-189 (1986)]}.

\bibitem[{\citenamefont{Baryshev et~al.}(1989)\citenamefont{Baryshev, Vasil'ev, Dmitriyev, Kupriyanov, Lukichev, Luk'yanova, and Sokolova}}]{vasilev1989}
\bibinfo{author}{\bibfnamefont{Y.~P.} \bibnamefont{Baryshev}}, \bibinfo{author}{\bibfnamefont{A.~G.} \bibnamefont{Vasil'ev}}, \bibinfo{author}{\bibfnamefont{A.~A.} \bibnamefont{Dmitriyev}}, \bibinfo{author}{\bibfnamefont{M.~Y.} \bibnamefont{Kupriyanov}}, \bibinfo{author}{\bibfnamefont{V.~F.} \bibnamefont{Lukichev}}, \bibinfo{author}{\bibfnamefont{I.~Y.} \bibnamefont{Luk'yanova}}, \bibnamefont{and} \bibinfo{author}{\bibfnamefont{I.~S.} \bibnamefont{Sokolova}}, \bibinfo{journal}{Lithography in microelectronics} \textbf{\bibinfo{volume}{8}}, \bibinfo{pages}{187} (\bibinfo{year}{1989}).

\bibitem[{\citenamefont{Bosboom et~al.}(2021)\citenamefont{Bosboom, Van~der Vegt, Kupriyanov, and Golubov}}]{bosboom2021}
\bibinfo{author}{\bibfnamefont{V.}~\bibnamefont{Bosboom}}, \bibinfo{author}{\bibfnamefont{J.~J.} \bibnamefont{Van~der Vegt}}, \bibinfo{author}{\bibfnamefont{M.~Y.} \bibnamefont{Kupriyanov}}, \bibnamefont{and} \bibinfo{author}{\bibfnamefont{A.~A.} \bibnamefont{Golubov}}, \bibinfo{journal}{Superconductor Science and Technology} \textbf{\bibinfo{volume}{34}}, \bibinfo{pages}{115022} (\bibinfo{year}{2021}).

\end{thebibliography}

\end{document}